\documentclass[acmsmall]{acmart}
\setcopyright{acmlicensed}
\acmJournal{PACMHCI}
\acmYear{2022} \acmVolume{6} \acmNumber{CSCW1}
\acmArticle{82} \acmMonth{4} \acmPrice{15.00}
\acmDOI{10.1145/3512929}

\AtBeginDocument{%
  \providecommand\BibTeX{{%
    \normalfont B\kern-0.5em{\scshape i\kern-0.25em b}\kern-0.8em\TeX}}}

\usepackage{graphics}
\usepackage{tikz}
\usepackage{subcaption}
\usepackage[normalem]{ulem}
\usepackage{tcolorbox}
\usetikzlibrary{shapes.geometric}

\begin{document}

\title[Comparing The Perceived Legitimacy of Content Moderation Processes]{Comparing the Perceived Legitimacy of Content Moderation Processes: Contractors, Algorithms, Expert Panels, and Digital Juries}

\author{Christina A. Pan}
\authornote{These authors contributed equally to this research.}
\email{cpan@cs.stanford.edu}
\orcid{0000-0002-0451-9796}
\affiliation{%
  \institution{Stanford University}
  \city{Stanford}
  \country{USA}
}

\author{Sahil Yakhmi}
\authornotemark[1]
\email{sahil@cs.stanford.edu}
\orcid{0000-0001-5120-0123}
\affiliation{%
  \institution{Stanford University}
  \city{Stanford}
  \country{USA}
}

\author{Tara P. Iyer}
\authornotemark[1]
\email{tiyer@alumni.stanford.edu}
\orcid{0000-0001-5666-2335}
\affiliation{%
  \institution{Stanford University}
  \city{Stanford}
  \country{USA}
}

\author{Evan Strasnick}
\email{estrasnick@stanford.com}
\orcid{0000-0002-1852-1854}
\affiliation{%
  \institution{Stanford University}
  \city{Stanford}
  \country{USA}
}

\author{Amy X. Zhang}
\email{axz@cs.uw.edu}
\orcid{0000-0001-9462-9835}
\affiliation{%
  \institution{University of Washington}
  \city{Seattle}
  \country{USA}
 }

\author{Michael S. Bernstein}
\email{msb@cs.stanford.edu}
\orcid{0000-0001-8020-9434}
\affiliation{%
  \institution{Stanford University}
  \city{Stanford}
  \country{USA}
}


\renewcommand{\shortauthors}{Christina A. Pan et al.}

\begin{abstract}

While research continues to investigate and improve the accuracy, fairness, and normative appropriateness of content moderation processes on large social media platforms, even the best process cannot be effective if users reject its authority as illegitimate. We present a survey experiment comparing the perceived institutional legitimacy of four popular content moderation processes. We conducted a within-subjects experiment in which we showed US Facebook users moderation decisions and randomized the description of whether those decisions were made by paid contractors, algorithms, expert panels, or juries of users. Prior work suggests that juries will have the highest perceived legitimacy due to the benefits of judicial independence and democratic representation. However, expert panels had greater perceived legitimacy than algorithms or juries. Moreover, outcome alignment---agreement with the decision---played a larger role than process in determining perceived legitimacy. These results suggest benefits to incorporating expert oversight in content moderation and underscore that any process will face legitimacy challenges derived from disagreement about outcomes.
\end{abstract}



\maketitle
\section{Introduction}

Efforts to improve platform design are ineffective if users do not trust platforms and their processes. Large social media platforms---like Facebook, YouTube, and Twitter---have become the new ``town squares'' for public discourse~\cite{le_roux_2020}, but the legitimacy of the rules and processes governing these platforms have increasingly been called into question. Social media platforms face widespread criticism for regulating speech in an opaque~\cite{roberts2019behind} and unrepresentative~\cite{kaye2019speech} manner without meaningful oversight~\cite{kaye2019speech}.

Since online platforms have become the ``new governors'' of speech~\cite{klonick2017new}, they have been analyzed through the lens of political theory and legitimacy~\cite{zuckerman2021mistrust, klonick2017new, suzor2019lawless, zhang_policykit, frey_ostrom, mou_2018, beer_2017}. Moreover, a large body of sociological work points to the practical importance of \textit{perceived legitimacy}---the acceptance of authority by those subject to it---for the functioning of institutions~\cite{tyler_2002, gibson_1992, mondak_1992, gibson_1998, hou_2017}. Empirical studies show that when institutions are perceived as highly legitimate by the public, this results in greater acceptance of unpopular decisions along with more cooperation and compliance with authorities in the long term. For instance, when the US Supreme Court decided a contentious election in Bush v. Gore, compliance was swift and the standing of the court was not measurably diminished~\cite{gibson2007legitimacy}. As institutions that regularly must make contentious decisions, online platforms similarly depend upon perceived legitimacy.

In this paper, we compare the perceived legitimacy of several content moderation processes that are in wide use or are specifically designed to increase legitimacy of moderation decisions. Understanding the impact of different content moderation processes on perceived legitimacy is critical---crafting even a ``perfect'' moderation process will not help a platform if that process is viewed by the population as illegitimate. In recent years, scholars have applied the lens of legitimacy to online platforms, including surveying the governance mechanisms they use~\cite{denardis_2015, gorwa_2019}, proposing frameworks with which to evaluate platform legitimacy~\cite{suzor_2018}, and proposing more legitimate methods of platform governance~\cite{santa_clara, kaye_smcs, hoover_verified_accountability, tworek2020dispute,fan2020digital}. However, most prior work lacks robust, empirical methods of evaluating legitimacy, and existing empirical work does not establish a common basis for comparing disparate processes. Stakeholders seeking to design more legitimate content moderation processes, whether platform owners, academics, or policymakers, currently lack data on how specific processes and proposals affect perceived legitimacy and the extent to which process design matters at all when making decisions about highly disagreed-upon content.


We conducted an online, within-subjects survey experiment in which US Facebook users evaluated moderation decisions presented as made by one of four processes: paid contractors, algorithms, expert panels, and juries of users. Paid contractors and algorithms are the two common types of content moderation used at scale~\cite{gillespie}, while expert panels like the Facebook Oversight Board~\cite{zuckFbOversightBoard} and digital juries~\cite{fan2020digital} are both recent moderation processes gathering substantial support and debate that are designed to enhance legitimacy.

In our within-subjects survey experiment, for each moderation process, participants were given a randomly selected Facebook post along with a randomly assigned decision outcome. For each post, participants were asked to answer questions about their attitudes towards the post and decision outcome, which measure components of perceived institutional legitimacy. At the end, participants were asked to compare and discuss the four processes. From participants' responses to the individual posts, we constructed a model that estimates the effect on perceived legitimacy of each moderation process, user alignment with the decision, and demographic variables. In addition to the quantitative analysis, we coded the comparative responses to identify and analyze all meaningfully distinguishable attitudes.

We find that expert panels have greater perceived legitimacy than both algorithms and digital juries. These results suggest that users value expertise, even when the nature of that expertise is not well understood. Additionally, we find qualitative evidence of a user preference for group decision making over decisions made by individuals and of acceptance of algorithmic decisions being conditional on factors like human oversight, despite being perceived as impartial. However, we also find that the alignment of user preferences with decision outcomes dominates all tested process factors in determining perceptions of legitimacy. In other words, whether users agree with the decisions of the content moderation process has a greater impact on the legitimacy users attach to that process than the process itself. While these results suggest platforms may struggle to create processes that can be perceived as legitimate by all users when dealing with highly disagreed-upon content, they also suggest incorporating expert oversight and multiple perspectives into moderation processes can help.

\section{Background And Research Questions}

In this section, we draw from prior work on platform governance, content moderation, and political legitimacy to motivate our study and methods. In contrast to the body of normative and qualitative work on content moderation, this paper contributes an empirical study that allows for user perceptions of multiple moderation processes to be compared with common methods of evaluating perceived legitimacy.

\subsection{Content Moderation Processes}

Each day, users posts billions of pieces of content on online platforms~\cite{gillespie}. This content must be reviewed so that illegal and harmful posts can be removed in a timely manner, while nevertheless respecting the users' right to self-expression~\cite{institute_montaigne, gillespie}. In this work, we consider the case of \textit{post hoc} content moderation takedown decisions, excluding processes involved in crafting content policy, to narrow the focus of the study to an intervention that is easily understood by participants and can be undertaken by several processes.

Online platforms employ varied methods to carry out this task. Consequently, researchers have sought to identify patterns in these strategies, for example, distinguishing between \textit{artisanal}, \textit{community-reliant}, and \textit{industrial} moderation processes~\cite{caplan2018content, gillespie}. Artisanal and community-reliant processes have been used by small platforms and niche communities within larger platforms like Reddit~\cite{gillespie, seering_2020}. However, the largest platforms heavily rely upon industrial moderation processes---defined as processes that enable platforms to 1) operate at large scale, 2) enforce well-defined rules, and 3) maintain separation between a) policy creation and b) interpretation and enforcement~\cite{caplan2018content}. Industrial moderation processes heavily overlap with commercial content moderation processes~\cite{roberts2019behind, gorwa_2020}.

In this work, we limit our scope to investigating the legitimacy of industrial moderation processes because they impact the most people, being employed by the largest platforms (Facebook, Youtube, Twitter, etc.), and are the processes most central to ongoing public debate over content moderation~\cite{caplan2018content}. As a result, we do not investigate community or artisanal moderation. Although community moderation is also employed by some large platforms, it is more closely linked to subcommunity norms rather than platform-wide rules, and generally does not strictly separate policy creation and enforcement. Community moderation, therefore, should be studied in context of specific community norms and not only from the perspective of \textit{post hoc} decisions.

We select the four following industrial moderation processes for our study.
\begin{enumerate}
    \item Paid Individual Contractors, who are hired and trained on a company's moderation policy
    \item Automated Systems, commonly powered by databases of known infringing content and machine learning algorithms, trained with the help of human contractors, that detect certain types of banned content (e.g., explicit language, hate speech, and pornographic images)
    \item Digital Juries, or ad-hoc deliberative bodies drawn from the user population
    \item Expert Panels, composed of experts in content moderation and related fields like law, human and digital rights, media and journalism, and political science
\end{enumerate}

Paid Individual Contractors (1) and Automated Systems (2) are selected because they are the industrial processes in widespread use by large platforms, and have been extensively described by many authors~(e.g.,~\cite{gillespie, caplan2018content, roberts2019behind, gorwa_2020}).

Digital Juries (3) and Expert Panels (4) are selected because they are emerging processes that also fall under the definition of industrial moderation. Both processes are well examined in the literature~\cite{fan2020digital, suzor, gill_2018}, are used in some form in industry~\cite{kou_2017, parler_community_jury, facebook_factchecking}, and were proposed specifically to help address the legitimacy issues plaguing earlier methods~\cite{democratic_legitimacy, suzor_2018, gill_2018, vaccaro2021contestability}. Digital Juries, as described by Fan and Zhang~\cite{fan2020digital}, draw legitimacy from democratic norms~\cite{democratic_legitimacy} and use of authentic deliberation~\cite{fan2020digital}. In industry, Digital Juries resemble juries as used on platforms such as League of Legends, Weibo, and even Parler~\cite{parler_community_jury,kou_2017}. Expert Panels are representative of bodies of experts like Facebook's fact-checking program using 3rd party fact-checkers~\cite{facebook_factchecking} or the Facebook Oversight Board~\cite{zuckFbOversightBoard}, which are intended to be transparent and independent. Because they are an emerging process, platforms are still developing the design of expert panels such as the Facebook Oversight Board, including how they can potentially scale and whether they should conduct policy creation separate from interpretation. However, we chose to include expert panels as a counterpoint to digital juries and limit it to be a process for policy interpretation in line with our other three processes.

\subsection{Legitimacy}
\label{sec:legitimacy}

\subsubsection{What is Legitimacy?}

\textit{Legitimacy} can be understood on either a normative or descriptive basis~\cite{stanford_legitimacy, jackson2018empirical}. In its normative sense, legitimacy “refers to some benchmark of acceptability or justification of political power or authority and—possibly—obligation”~\cite{stanford_legitimacy}. Influential examples of normative legitimacy frameworks include constitutional legitimacy~\cite{rosenfeld2000rule} and democratic legitimacy~\cite{peter2009democratic}, discussed later in the context of emerging content moderation processes. By contrast, in its descriptive sense, legitimacy refers to the acceptance of authority~\cite{stanford_legitimacy, weber1964theory}.

This study examines legitimacy in its descriptive sense, i.e., as a measurable, subjective, sociological phenomenon, referred to using the more common phrase \textit{perceived legitimacy}~\cite{tyler2007legitimacy}. However, while we emphasize perceived legitimacy for its established practical benefits, we recognize the role that normative principles like fairness play in shaping attitudes~\cite{jackson2018empirical}. As such, we will discuss prior work examining both conceptions of legitimacy, using \textit{legitimacy} to refer to the expansive concept in both its normative and descriptive senses. We use the term \textit{democratic legitimacy} to refer to the normative concept that political systems derive legitimacy through adherence to democratic norms, procedures, and values~\mbox{\cite{democratic_legitimacy}}.

\subsubsection{Measuring Perceived Legitimacy}

Modern social scientists have contributed a wealth of work on measuring the perceived legitimacy of governance. We draw primarily from work studying the perceived legitimacy of the courts. Among the best established of this work is that of law and psychology professor Tom Tyler and political scientist James Gibson. In Tyler’s framework, fair procedure, quality of decision making, quality of treatment, and motive-based trust contribute to greater perceived legitimacy, while perceived legitimacy in turn fosters compliance, cooperation, and empowerment~\cite{tyler2003procedural}. Tyler also highlights that, to a plurality, perceived legitimacy is analogous to obtaining the person's desired outcome~\cite{tyler_youtube}, implying that there is a limited extent to which process design can create perceived legitimacy at all. While Tyler examines individuals' interactions with the state, Gibson instead frames perceived legitimacy around institutions. Gibson measures the legitimacy of institutions like courts through procedural values like trustworthiness and neutrality. In addition, he measures institutional commitment, the extent to which people support an institution’s existence, and decisional jurisdiction, the support for the institution’s power over a particular application~\cite{gibson2003defenders}. Gibson places special emphasis on ``diffuse support,'' a ``reservoir of favorable attitudes or good will that helps members to accept or tolerate outputs to which they are opposed or the effects of which they see as damaging to their wants,'' as opposed to ``specific support,'' or support for a particular action or policy~\cite{easton1965systems, gibson_1992}.

In the domain of content moderation, however, little work exists that measures perceived legitimacy. Instead, most prior work investigates questions of normative legitimacy, for example outlining fundamental rights and procedural values known to correspond to legitimate governance~\cite{suzor_2018, suzor_2019_transparency}. Of the studies that take a more descriptive and empirical approach, none have tackled the question of perceived legitimacy head on, focusing instead on adjacent questions~\cite{shoenebeck_2018, fan2020digital, vaccaro_2020}.

In the absence of an established measure of perceived legitimacy of content moderation, we select Gibson's formulation of institutional legitimacy as our overarching framework for measuring perceived legitimacy. In addition, we follow Gibson in using population-wide measures of attitudes to capture ``diffuse support.''

\subsection{Studying the Legitimacy of Content Moderation Processes}

From prior work, we can conclude that while legitimacy is broadly accepted as an important and desirable quality in content moderation systems and a variety of perspectives exist regarding how it can be accrued, the impact of specific processes on perceived legitimacy remains largely unknown.

This missing data in the literature motivates the primary research question of our study:

\begin{quote}
    \begin{tcolorbox}
    \textbf{\textit{RQ1:} How do content moderation processes impact levels of perceived institutional legitimacy?}
    \end{tcolorbox}
\end{quote}

For those seeking to design legitimate content moderation systems, a major open question is the extent to which process design can create perceived legitimacy at all. Because a legitimate process is most valuable when it can mitigate the negative effects of an unfavorable decision, it is important to contextualize the magnitude of process effects by comparing them with the strength of outcome effects. Thus we also ask:

\begin{quote}
    \begin{tcolorbox}
    \textbf{\textit{RQ2:} To what extent does the alignment of outcome of decisions made by a process with individual preferences influence the perceived institutional legitimacy of that process?}
    \end{tcolorbox}
\end{quote}

\subsection{Known Determiners of Legitimacy}

Academics have proposed a plethora of principles and frameworks that contribute to legitimate governance, including transparency and public participation~\cite{suzor_2019_transparency, grimmelmann2015virtues, fung_2013, hood_2012, mcintyre2008internet}, adherence to established legal principles~\cite{klonick2017new, balkin2015information}, and upholding individual rights~\cite{kaye2018human, citron2009cyber}. However, to develop hypotheses for \textbf{RQ1}, we focus our discussion on prior work examining the factors that differ between the four processes.

\subsubsection{Independence}

Among a large space of dispute resolution processes, prior work shows a consistent preference among litigants for greater decision control by an impartial third-party~\cite{houlden_1978, shestowsky_2004}. National high courts enjoy special legitimacy~\cite{gibson_1998}, and it is often taken for granted that their greater independence contributes to public trust~\cite{levasseur_2002}. Multiple empirical studies also find support for independence conferring greater perceived legitimacy to political institutions and courts~\cite{gibson_2008, buhlmann_2011}. Consequently, in the domain of content moderation, decision-making by independent bodies commonly features in high level frameworks designed to enhance legitimacy, such as FAITE~\cite{tworek2020dispute}, national social media councils (SMCs)~\cite{kaye_smcs}, and policy proposals by the Cato Institute and the Bookings Institute~\cite{cato_legitimacy, brookings_zuckerberg}.

Among the four moderation processes in this study, we consider the digital jury to have high independence by analogy with criminal juries, which serve as an independent check on government power~\cite{solomon2011political}. Additionally, we expect jury members' loose and impersonal relationship with the platform would limit the platform's influence over their decisions. We accept the judgment of prior work that expert panels can benefit from independence~\cite{SLSpolicylabreport}, but note that potential platform influence over the body and its composition may limit practical independence. Conversely, we deem human contractors and algorithmic moderation to have no practical independence from the platform.

Due to the greater independence of the deliberative bodies, we hypothesize:

\begin{quote}
\begin{tcolorbox}[standard jigsaw, opacityback=0]
\textbf{H1.1:} The expert panel and digital jury will be perceived as more legitimate than the paid contractor and algorithm.
\end{tcolorbox}
\end{quote}

\subsubsection{Automated Decision Making}

As the use of Algorithmic Decision Making (ADS) has grown, researchers have studied its characteristics relative to human decision making from multiple perspectives~\cite{castelluccia2019understanding}. ADS is often evaluated according to to specific normative criteria, including fairness~\cite{barocas-hardt-narayanan, hutchinson_2019}, accountability~\cite{wieringa}, explainability~\cite{adadi2018peeking, mittelstadt_2019}, and contestability~\cite{hirsch2017designing, vaccaro_2020}. Such inquiry is motivated in part by evidence that ADS can be biased and can cause various types of harms~\cite{barocas2016big, obermeyer_2019, barabas_2017, crawford2017trouble}, and in part by application of theories of justice~\cite{lundgard_2020, jurgens-etal-2019-just, 10.1145/3173574.3173951}. Despite major theoretical and practical issues commonly known in  academia, prior work shows that perception of the trustworthiness of algorithms relative to humans can be favorable, though it is highly dependent on context, subjectivity of the domain, and performance~\cite{araujo_2020, lee_2018, logg}. Moreover, public perceptions of algorithms are subject to cognitive biases, including overconfidence in their capabilities~\cite{waggoner2019big}, outcome favorability bias~\cite{wang_2020, eslami_2019}, excessive aversion to mistakes~\cite{dietvorst2015algorithm}, and folk theories~\cite{eslami_2016}. In general, algorithms tend to benefit from being perceived as impartial, objective, and authoritative by the public~\cite{Gillespie2013TheRO, sundar2008main}.

In this study, algorithmic moderation embodies automated judgment, standing in contrast to deliberative bodies like the expert panel and digital jury. While paid contractors can exercise human judgment, we assess that because paid contractors are given extremely limited time and detailed guidelines to make decisions~\cite{gillespie}, their decisions involve significantly less discretion.

For highly disagreed-upon posts, we anticipate that perceptions of algorithms' impartiality will outweigh concerns about their lack of ability in a subjective domain and the larger penalties they receive for poor performance. Thus, making a direct comparison between the automated and human processes with low independence, we hypothesize:

\begin{quote}
\begin{tcolorbox}[standard jigsaw, opacityback=0]
\textbf{H1.2:} The algorithm will be perceived as more legitimate than the paid contractor.
\end{tcolorbox}
\end{quote}

\subsubsection{Democratic Legitimacy vs. Expertise}

The debate over judge vs. jury trials in the judicial system can be understood as a debate over democratic vs. expert authority. Juries, despite well known drawbacks~\cite{klein2016truth}, have long been justified on the grounds that they bind the legal system to community norms and provide legitimacy through democratic representation and the exercise of popular sovereignty~\cite{farrar2001echoes, solomon2011political, schwartzbergcivil}. Surveys in the US have consistently found broad public support for juries as an institution (``diffuse support'')~\cite{hans2019trial}. However, little rigorous empirical work exists that compares the perceived legitimacy of juries to judges~\cite{solomon2011political, klein2016truth}. Nevertheless, surveys tend to show a preference for juries over judges~\cite{solomon2011political} and some empirical evidence of perceived legitimacy benefits of citizen participation have been noted in multiple countries~\cite{machura2003fairness, bergoglio2017ten}.

Prior work on the perceived legitimacy of expert authority is mixed. On the one hand, public trust in experts appears pervasive~\cite{pew_trust, lse_experts}, and expertise has traditionally been seen as a way to establish legitimate authority~\cite{weber2009theory, french1959}. However, critiques of expert authority are common, with scholars pointing to unequal relationships between experts and the public and other issues~\cite{turner2001problem, habermas1984theory, foucault1980power}. Steven Turner resolves this tension by noting that claims to cognitive authority must be legitimated through acceptance by the public, observing that across fields experts achieve varying levels of success~\cite{turner2001problem}.

In this study, the expert panel embodies expert knowledge and judgment, while the digital jury represents democratic participation. Contractors and algorithms, though they may act in accordance with expert-designed guidelines, do not exercise sufficient individual discretion to represent either type of knowledge.

Prior work generally supports the idea that juries are perceived as more legitimate decision makers, and suggests that the benefits of democratic legitimacy extend to the domain of content moderation~\cite{fan2020digital}. Moreover, because the domain of content moderation is relatively novel, it is reasonable to expect that the legitimation process of experts among the public---which can lag legitimation among professionals by decades~\cite{turner2001problem}---is still in its infancy. Consequently, we hypothesize:

\begin{quote}
\begin{tcolorbox}[standard jigsaw, opacityback=0]
\textbf{H1.3:} The digital jury will be perceived as more legitimate than the expert panel.
\end{tcolorbox}
\end{quote}

In the following subsection, we review prior work relating to \textbf{RQ2}.

\subsubsection{Role of Pre-Existing Views}

While much work on perceived legitimacy is concerned with its ability to promote the acceptance of adverse or unpopular outcomes, prior work suggests that perceived legitimacy is itself shaped by alignment with individual preferences and beliefs. A large body of work finds evidence for various confirmation or congeniality biases~\cite{KLAYMAN1995385, bohner_attitude_change} whereby pre-existing views affect how information is collected~\cite{klapper1960effects}, interpreted~\cite{lord_1979}, and evaluated~\cite{druckman2011framing}. Similarly, motivated reasoning theory describes mechanisms by which directional goals bias cognitive processes~\cite{kunda1990case}. These biases can be strongly mediated by partisan identification and cues~\cite{leeper2014political, goren2009source}. There is some indication that these cognitive effects may extend to perception of legitimacy. While Gibson finds that controversial decisions do not necessarily impair the legitimacy of an institution like the Supreme Court~\cite{gibson2007legitimacy}, other work finds that strongly held moral convictions do magnify the effect of outcomes on perceived legitimacy of the court~\cite{skitka2009limits}.

Based on this prior work, we hypothesize:

\begin{quote}
\begin{tcolorbox}[standard jigsaw, opacityback=0]
\textbf{H2:} Users will report higher perceived legitimacy when content moderation systems make decisions that align with their individual normative preferences about whether a post should stay up or be taken down, and this association will be at least as strong as that of process factors.
\end{tcolorbox}
\end{quote}

\section{Methods}

This study records and analyzes how US Facebook users perceive the institutional legitimacy of various content moderation processes in an online survey setting. In contrast to prior work~\cite{jhaver_2019, shoenebeck_2018}, we examine the attitudes of users who are not directly involved in content takedown decisions (i.e., bystanders). The literature on legitimacy indicates that public attitudes (i.e., ``diffuse support'') are what determine the legitimacy of institutions~\cite{gibson_1992}. Moreover, in online communities, the vast majority of users are never involved in content moderation disputes~\cite{nielsen_2006}. We designed the survey around Facebook due to the platform's scale~\cite{statista_fb_users}, broad adoption in the USA~\cite{statista_fb_age_usa}, and representative user base~\cite{fb_diversity_us}, and we recruited participants from Amazon Mechanical Turk (AMT), following a common practice in political science studies~\cite{coppock2019generalizing}.

\subsection{Materials}

We followed the example set by prior work in collecting real social media posts rather than synthesizing controlled examples~\cite{fan2020digital}. This approach mitigates potential biases in post creation and improves the ecological validity of the study, as prior work suggests that hypothetical choices can differ from choices made in concrete situations~\cite{kuhberger2002framing}. To reduce the impact of biases in post selection, we employed a two-stage strategy, described below.

We first compiled a list of Facebook posts representing a wide array of topics common in takedown decisions (e.g., racism, protest, vaccination, electoral fraud, government conspiracy) and viewpoints (e.g., both liberal and conservative), taking care to avoid specific posts that participants were likely to have already encountered in the media. We collected posts that might be viewed as violating Facebook's Community Standards~\cite{fb_community_standards} in three of its categories: inciting violence, hate speech, and misinformation. These categories were chosen for their high frequency and prominence in public disagreements about content moderation. We collected 58 candidate posts from three sources: public Facebook groups, low-traffic news articles, and the Plain View Project (PVP)~\cite{plain_view_project}. The PVP is a journalistic database of Facebook posts authored by police officers expressing themes of violence, racism, and bigotry. To find Facebook groups and news articles, we identified common topics within each category that elicited public disagreement (e.g., anti-vaccination in misinformation). We then used these topics as search terms on Facebook to find groups and on news search engines to find articles containing posts. The resulting posts may or may not have actually been removed by Facebook.

To further mitigate bias, we narrowed this broad pool of candidates to nine posts (i.e., 3 in each category of potential infringement) by selecting the posts that were the most disagreed-upon by Facebook user participants on AMT. We ran a pre-survey that asked 56 participants (not eligible for the main survey) their opinion about whether a given post ought to be removed (i.e., normative preference).  Responses were recorded on a five-point Likert scale, ranging from strongly disagree (1) to strongly agree (5). For each candidate post, we calculated its \textit{disagreement score} as a combination of the standard deviation of the responses and the absolute deviation of the median from the neutral response value of 3: $disagreement_{post} = \sigma_{post} -|3-\mu_{post}|$. This formulation was chosen to ensure that posts would not only elicit a wide spread of opinions, but these opinions would be well balanced between favoring taking down and leaving up.\footnote{ After the study was conducted, we discovered that we mistakenly included a post (Post 5 in the Supplementary Materials) that was not among the top 3 posts by disagreement score in its category---inciting violence. Although the study was designed around highly disagreed-upon content to enhance our ability to measure effects of process on perceived legitimacy, content moderation processes also deal with content for which opinions are more homogeneous. After performing additional analysis on a dataset that excluded the post in question, we found no meaningful change in the magnitude or direction of effects but observed higher p-values due to the loss of about 11\% of data.} The median standard deviation of responses for the final nine posts was $1.5$, and the median of the posts' median response was $3.0$. These posts are available in the Supplementary Materials.

\subsection{Experimental Design}

We constructed a within-subjects survey experiment to assess the perceived institutional legitimacy of content moderation decisions made by a paid contractor, an algorithm, an expert panel, and a digital jury. Participants were given 4 randomly constructed content moderation decisions---this randomization exposed participants to many many combinations of posts and processes to help mitigate biases introduced by individual posts. For each decision, participants were asked to answer several questions regarding their attitudes toward the post and the decision outcome. At the end, participants were asked to discuss the four processes on a comparative basis. The study design was reviewed and approved by our institution's institutional review board (IRB) under protocol \#57848. Selected screenshots of the survey are available in supplementary materials.

\subsubsection{Participants}

The survey was sent to US Facebook users on AMT. AMT allows only workers 18 years or older, and gives workers the option to self-report being Facebook users.  Participants were required to go through an IRB-approved consent process with appropriate content warnings and resources. Participants were informed that neither the moderation decisions nor processes were real only in the survey debrief, to improve the survey realism. Participants were compensated \$1.82 for the 15 minutes spent completing the survey, based on the 2020 federal minimum wage of \$7.25/hr~\cite{fed_min_wage}, a rate above the mean and median hourly wages for AMT workers (\$3.13/hr and \$1.77/hr, respectively)~\cite{hara2018data}.

We set a target sample size of 100 participants based on a small pilot study in which we already observed significant outcome-preference alignment effects, and power analysis aiming to detect an effect size of 0.5 points (out of 20) for process effect. After the data validation described below, 93 responses remained. No additional stratified (sub)sampling was performed.

Participants were 57\% female and 43\% male. Participants were also well balanced between political affiliations, with 35\% identifying as liberal, 31\% as conservative, 31\% as independent, and a remaining 2\% refraining from reporting affiliation. Roughly 60\% of participants were between the ages of 25 and 44, with 35\% 45 or older. This age distribution mirrors that of US Facebook users~\cite{statista_fb_age_usa}, although it underrepresents the 18-24 age group. The survey population reported as 80\% White, 11\% Asian, 4\% Mixed Race, 3\% Black, and 1\% Native American, with a further 1\% declining to report. Additionally, 11\% of participants reported as Hispanic or Latino, across all race categories. Compared to the US population, our survey population was more educated, with only 23\% reporting highest attainment as high school, 38\% with a Bachelor's degree, and 18\% with a Master's or higher.
\subsubsection{Experimental Manipulation} Each participant was shown four moderation decisions consisting of 1) a post randomly selected from the nine, 2) one of the four moderation processes, 3) a random decision outcome---taken down or left up, and 4) a brief indication of the violation category if the post was taken down. Each moderation process was shown exactly once, in random order, and posts were sampled such that each of the three categories of content violation would be seen at least once in the four decisions. The moderation process descriptions shown to participants were intentionally kept short to allow pre-existing attitudes and assumptions to be captured in responses, and to approximate the opaque nature of content moderation as practiced today~\cite{roberts2019behind}. These descriptions are provided in Appendix A. From our pilot studies, we found that the descriptions were adequate for users to be able to understand and differentiate between the moderation processes, aligning with prior work~\cite{logg, araujo_2020, lee_2018}.

\subsubsection{Measures} \label{sec:measures} 
\begin{table}[tb]\centering
\begin{tabular*}{\columnwidth}{@{\extracolsep{\fill}} lp{8cm} }
\toprule
\textit{Measure of Institutional Legitimacy }   & \textit{Question} \\ 
\midrule
Outcome Satisfaction           &      I am satisfied with the way the [moderation process] handled this moderation decision. \\
Trustworthiness           &     [Moderation process] can be trusted.     \\
Fairness and Impartiality           &      [Moderation process] can be fair and impartial.     \\
Institutional Commitment           &      Facebook should keep using [moderation process] to make content moderation decisions.     \\
Decisional Jurisdiction           &      [Moderation process] should be the authority making moderation decisions.     \\
\bottomrule
\end{tabular*}
\caption{Questions used to measure each aspect of institutional legitimacy in the survey experiment. Users responded to these questions on a 5 point Likert scale. [Moderation process] should be replaced with the moderation process in question (e.g., digital juries of Facebook users).}
\label{table:measures}
\end{table}
\pagebreak
The perceived institutional legitimacy of moderation decisions served as the primary quantitative measure of the survey. Five survey questions, given in Table \ref{table:measures}, were posed to participants for each moderation decision, corresponding to five component measures of perceived institutional legitimacy---outcome satisfaction, users' trust in the process, perceived fairness and impartiality, institutional commitment, and decisional jurisdiction. The questions assessing trustworthiness, institutional commitment, and decisional jurisdiction were adapted from Gibson's work surveying the institutional legitimacy of national high courts~\cite{gibson_1998, gibson2003defenders}, while the question assessing fairness and impartiality was adapted from a study measuring perceived legitimacy of state Supreme Courts~\cite{gibson_2008}. These questions were modified to fit the domain, and institutional commitment and decisional jurisdiction were flipped from negative to affirmative to better suit our hypothetical setting. We also included a question assessing outcome satisfaction (found to be positively correlated with institutional legitimacy~\cite{gibson2003defenders}) using similar language to prior work evaluating content moderation~\cite{fan2020digital}. As in prior empirical work~\cite{gibson_1998, gibson2003defenders, fan2020digital}, terms like fairness and impartiality were not rigorously defined to avoid unduly influencing participants with prescriptive normative criteria. The responses to these questions were captured on a five point Likert scale, and we calculated Cronbach's alpha (a common measure of internal consistency) between these five measures in our data as 0.92. These component measures were summed to create a composite measure.

In addition to this quantitative measure, the survey also collected qualitative data through free response questions. Participants were randomly asked to elaborate on their responses to quantitative questions 50\% of the time. Additionally, after answering questions about the four moderation decisions, all participants were asked 1) to select the process they saw as the most trusted, least trusted, most fair and impartial, and least fair and impartial, and 2) to provide a brief rationale(s) behind their choices. These comparative questions were included not only to corroborate quantitative results, but also because prior research shows that people can be more effective in making comparative judgments~\cite{yannakakis_2015}.

Demographic information, including age, gender, race, ethnicity, education level, work experience, political affiliation, income, and Facebook usage, was also collected primarily to assess the representativeness of the participant group, and in limited cases to test for association with perceived legitimacy (detailed below).

\subsubsection{Data Validation}

In order to validate responses, users were asked to answer attention check questions (repeating back details about the moderation process and outcome). Any participants that failed these attention check questions were removed from the dataset. In addition, any spam-like submissions were removed.

\subsubsection{Quantitative Modeling}

Quantitative responses were analyzed using a linear mixed effects (LME) model in which the degree of alignment of individual normative preference with outcome (\textsc{Alignment}), content moderation process (\textsc{Process}), \textsc{Gender}, and \textsc{Political Affiliation} serve as explanatory variables, and measures of perceived legitimacy, as the response variable (as described in Section \ref{sec:measures}). The inclusion of \textsc{Alignment} and \textsc{Process} relate to \textbf{RQ2} and \textbf{RQ1} respectively, while \textsc{Gender} and \textsc{Political Affiliation} are included because they have been shown to relate to perceived legitimacy in prior work~\cite{tyler2003procedural, gibson2014}. We intentionally do not control for participants' prior exposure to content moderation, as perceived legitimacy measures population-wide attitudes---it is a sociological phenomenon that must be assessed within a representative population sample. In this model, each participant is given a random intercept and slope for decision outcome, allowing for the possibility that each participant may have a different inclination to take down or leave up posts. Additionally, each post is given a random intercept and slope for decision outcome and political affiliation, as specific posts may be more or less objectionable across the population, and many posts have a significant political dimension.

The composite measure of perceived legitimacy serves as the dependent variable for the primary model, which is used for all hypothesis tests. To further understand how the explanatory variables relate to individual measures of perceived legitimacy, parallel submodels were also fit with each of the five perceived legitimacy measures as dependent variables.

\subsubsection{Qualitative Coding}
To analyze the four final comparative free response questions, two co-authors identified the moderation process named by each participant. If no process could be identified, the entire response was excluded from analysis. If multiple processes were indicated, only the process identified as a first choice was coded if a relative ordering was given, otherwise all processes were coded. The same two co-authors then performed an open coding procedure to identify all meaningfully distinguishable attitudes expressed by participants in their answers. Noting that participants frequently hedged their answers and expressed multiple attitudes at a time, one co-author then developed a framework of axial codes in which attitudes were coded as a series of $(role, subject, predicate)$ triples, each component of which is defined as follows:
\begin{enumerate}
    \item \textit{role}: Indicates whether the attitude served as a \textit{rationale} for the answer, \textit{qualification} of the answer, or \textit{condition} for the answer.
    \item \textit{subject}: Indicates to which of the four moderation process(es) the attitude pertains.
    \item \textit{predicate}: Indicates the idea being expressed about the subject.
\end{enumerate}

For example, the attitude expressed in, ``I think a panel of experts can be most trusted because they have the training needed to make good decisions, but the platform can select experts in a biased way,'' might be coded as: $[(rationale,\allowbreak Expert,\allowbreak \text{Code 8}: \textit{"has necessary training"}),\allowbreak (qualification,\allowbreak Expert,\allowbreak \text{Code 13}: \textit{"controlled by platform"})]$. The two co-authors independently rated all responses, and Cohen's kappa, a metric of inter-rater reliability, was calculated separately for each possible code. Across processes we calculated a mean kappa of 0.99, and across attitude triples we calculated a frequency-weighted average kappa of 0.61. Subsequently, the two co-authors discussed inconsistencies and reached unanimous agreement on the final coding of each response. The full attitude coding scheme contains approximately 50 distinct predicates, which are given in Appendix B.

\section{Results}

\begin{table}[tb]\centering
\begin{tabular*}{\columnwidth}{@{\extracolsep{\fill}} lll }
\toprule
\textit{Variable}   & \textit{Alternative: No Interactions} & \textit{Primary Model} \\ 
\midrule
Alignment           &\textbf{1.86***} &\textbf{1.87***}\\
                    &      (0.15)     &      (0.15)   \\
Algorithm           &      -0.66      &\textbf{-0.68*}    \\
                    &      (0.34)     &      (0.34)   \\
Expert Panel        &\textbf{1.18***} &\textbf{1.14***}    \\
                    &      (0.34)     &      (0.34)   \\
Digital Jury        &      -0.56      &      -0.55    \\
                    &      (0.34)     &      (0.34)   \\
Male                &      -0.52      &      -0.45    \\
                    &      (0.67)     &      (0.67)   \\
Conservative        &      -1.04      &      -1.10    \\
                    &      (0.84)     &      (0.84)   \\
Independent         &\textbf{-2.14*}  &\textbf{-2.18**}\\
                    &     (0.83)      &      (0.82)   \\
Unreported Affiliation&   -3.80       &      -3.78    \\
                    &     (2.32)      &      (2.32)   \\
Alignment * Algorithm&                &       0.06    \\
                    &                 &      (0.26)   \\
Alignment * Expert  &             &       0.15    \\
                    &                 &      (0.25)   \\
Alignment * Jury&   &       0.17    \\
                    &                 &      (0.24)   \\
Constant            &\textbf{17.68***} &\textbf{17.71***}   \\
                    &      (0.64)     &      (0.65)   \\
\bottomrule
\end{tabular*}
\caption{Determinants of perceived legitimacy as modeled by primary and alternative LME models. Process is modeled using deviation contrasts such that the Constant reflects the mean across processes, and coefficients are comparable between models. Significance levels are indicated for readability purposes only and are calculated with t-tests using Satterthwaite's method~\cite{luke2017evaluating}. These levels do not constitute formal hypothesis tests. | $p<0.001$\textsuperscript{***}, 
  $p<0.01$\textsuperscript{**}, 
  $p<0.05$\textsuperscript{*}}
 \vspace{-7mm}
\label{table:regression}
\end{table}

The primary quantitative model of survey responses estimates the effect on perceived institutional legitimacy of \textsc{Alignment}, \textsc{Process}, \textsc{Gender}, and \textsc{Political Affiliation}. Regression coefficients from this model are presented in Table \ref{table:regression}, where coefficients from an alternative model without interaction terms are also given for comparison. While statistical tests do not show greater explanatory power for the primary model versus this alternative, the full model is used for hypothesis tests. Results from the parallel submodels are found to be consistent with the primary model---suggesting that the composite measure is not dominated by a subset of measures. Regression results from the submodels are given in Appendix C.

The effects of \textsc{Process} and \textsc{Alignment} are discussed in detail below. We do not find evidence that \textsc{Gender} is associated with perceived legitimacy. We do find evidence that political affiliation has a statistically significant relationship with perceived legitimacy by ANOVA, but pairwise contrasts are not statistically significant, preventing us from drawing specific conclusions.

\begin{table}[tb]
\vspace{0pt}
\begin{tabular}{lcccc} \toprule
                \textit{Process} & \multicolumn{4}{c}{\textit{Proportion of Respondents (\%)}}                                                                                                    \\
\midrule
                & \multicolumn{2}{c}{\textit{Trustworthiness}}                              & \multicolumn{2}{c}{\textit{Impartiality}}                                  \\
\cmidrule(r){2-3}
\cmidrule(r){4-5}
& \multicolumn{1}{c}{\textit{Highest}} & \multicolumn{1}{c}{\textit{Lowest}} & \multicolumn{1}{c}{\textit{Highest}} & \multicolumn{1}{c}{\textit{Lowest}} \\
\midrule
Contractor & 14\% {\small (13)}&\textbf{35\%} {\small\textbf{(31)}}& 9\% {\small (8)}&\textbf{46\%} {\small \textbf{(41)}} \\
Algorithm & 30\% {\small (28)}& 34\% {\small (30)}&\textbf{51\%} {\small \textbf{(46)}}& 13\% {\small (12)}\\
Expert &\textbf{41\%} {\small \textbf{(38)}}& 9\% {\small (8)} & 28\% {\small (25)}& 8\% {\small (7)}\\
Jury &   28\% {\small (26)}& 27\% {\small (24)}& 18\% {\small (16)}& 36\% {\small (32)}\\
\bottomrule
\end{tabular}
\caption{Proportion of participants indicating each moderation process as ranking the highest and lowest with respect to \textit{trustworthiness} and \textit{impartiality}. Raw participant counts are given in parentheses.}
\label{table:qual_methods}
\end{table}

\subsection{Perceived Legitimacy of Moderation Processes}

\begin{table}[tb]\centering
\begin{tabular}{lll}
\toprule
\textit{Contrast}   & \textit{Estimate} & \textit{Standard Error} \\ 
\midrule
Algorithm - Contractor      &      -0.78    &   0.60   \\
Expert - Contractor   &         1.02    &   0.60   \\
Expert - Algorithm    & \textbf{1.81*}  &   0.60   \\
Jury - Contractor   &        -0.66    &   0.60   \\
Jury - Algorithm    &         0.13    &   0.62   \\
Jury - Expert &\textbf{-1.68*}  &   0.61   \\
\bottomrule
\end{tabular}
\caption{Tukey's HSD test results of significance in the difference in mean perceived legitimacy across moderation processes. | $p<0.001$\textsuperscript{***}, 
  $p<0.01$\textsuperscript{**}, 
  $p<0.05$\textsuperscript{*}}
 \vspace{-7mm}
\label{table:h1}
\end{table}

\begin{figure}
    \centering
    \includegraphics[scale=0.5]{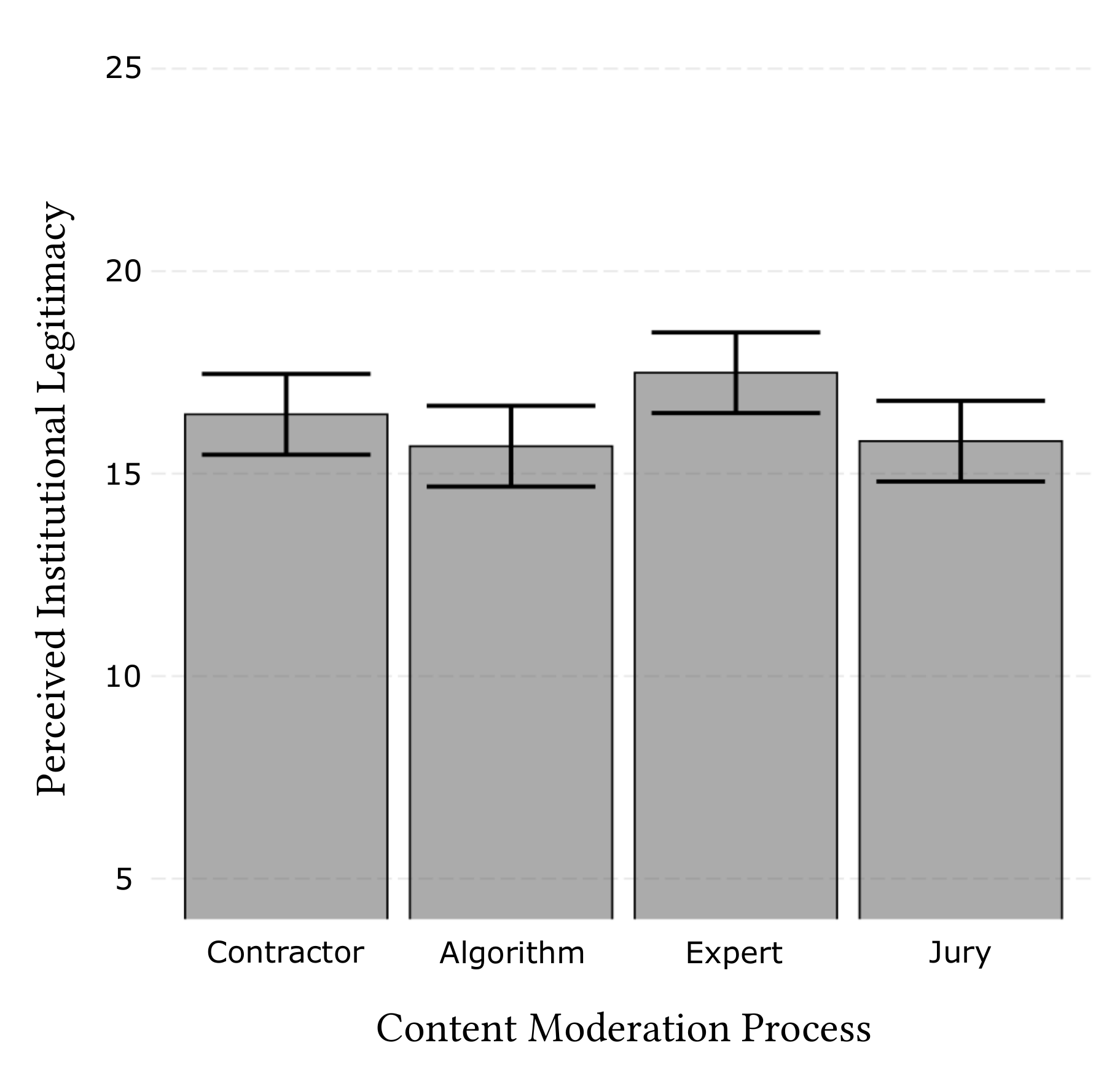}
    \caption{The mean perceived legitimacy of each process is plotted, shown with a 95\% confidence interval calculated from variance of the fixed effect estimates.}
    \label{fig:leg_v_method}
\end{figure}

\begin{figure}
\centering
\begin{subfigure}{.5\textwidth}
  \centering
  \includegraphics[width=0.85\linewidth]{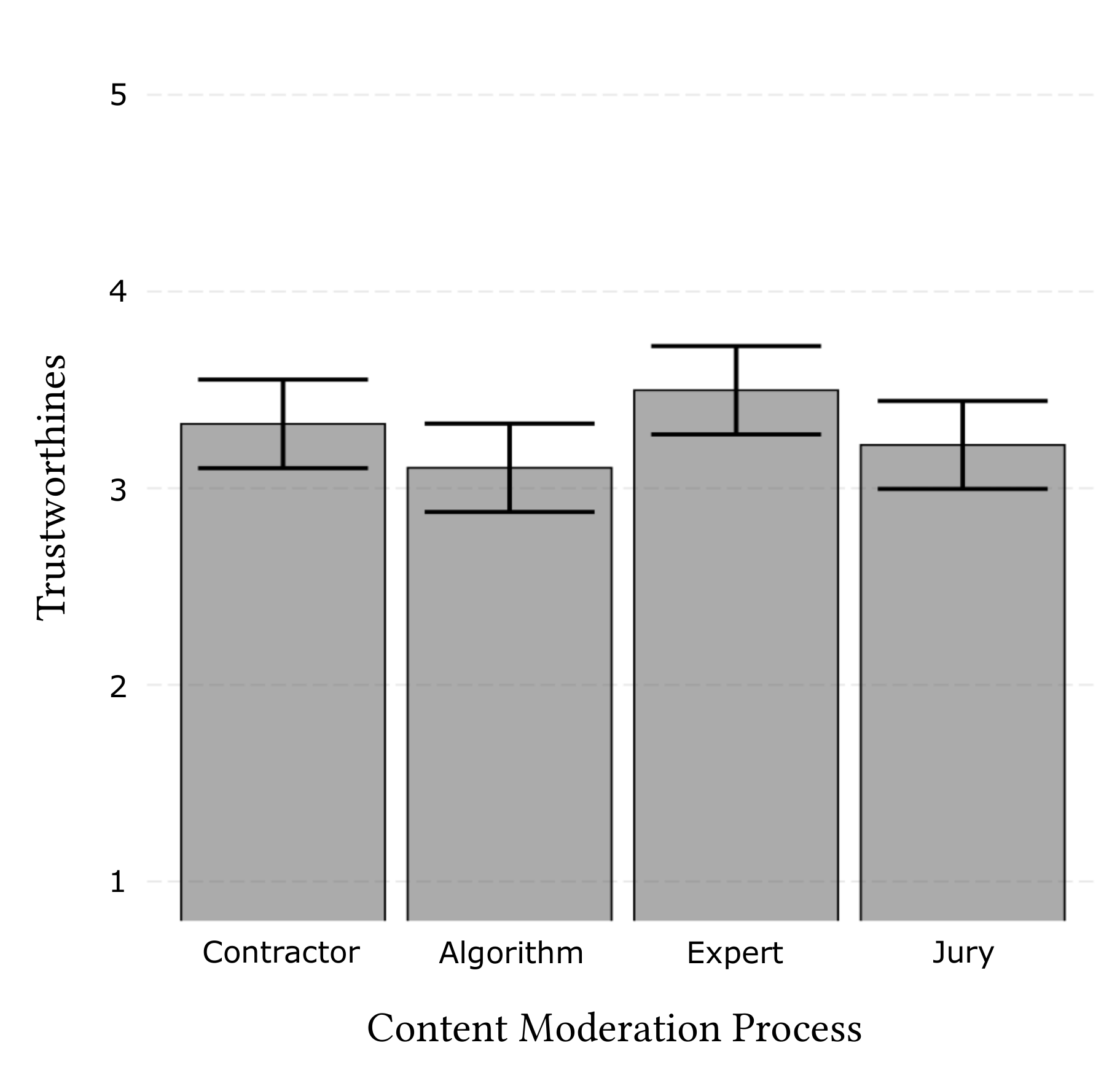}
  \caption{The mean \textit{trustworthiness} of each process.}
  \label{fig:trust_v_method}
\end{subfigure}%
\begin{subfigure}{.5\textwidth}
  \centering
  \includegraphics[width=0.85\linewidth]{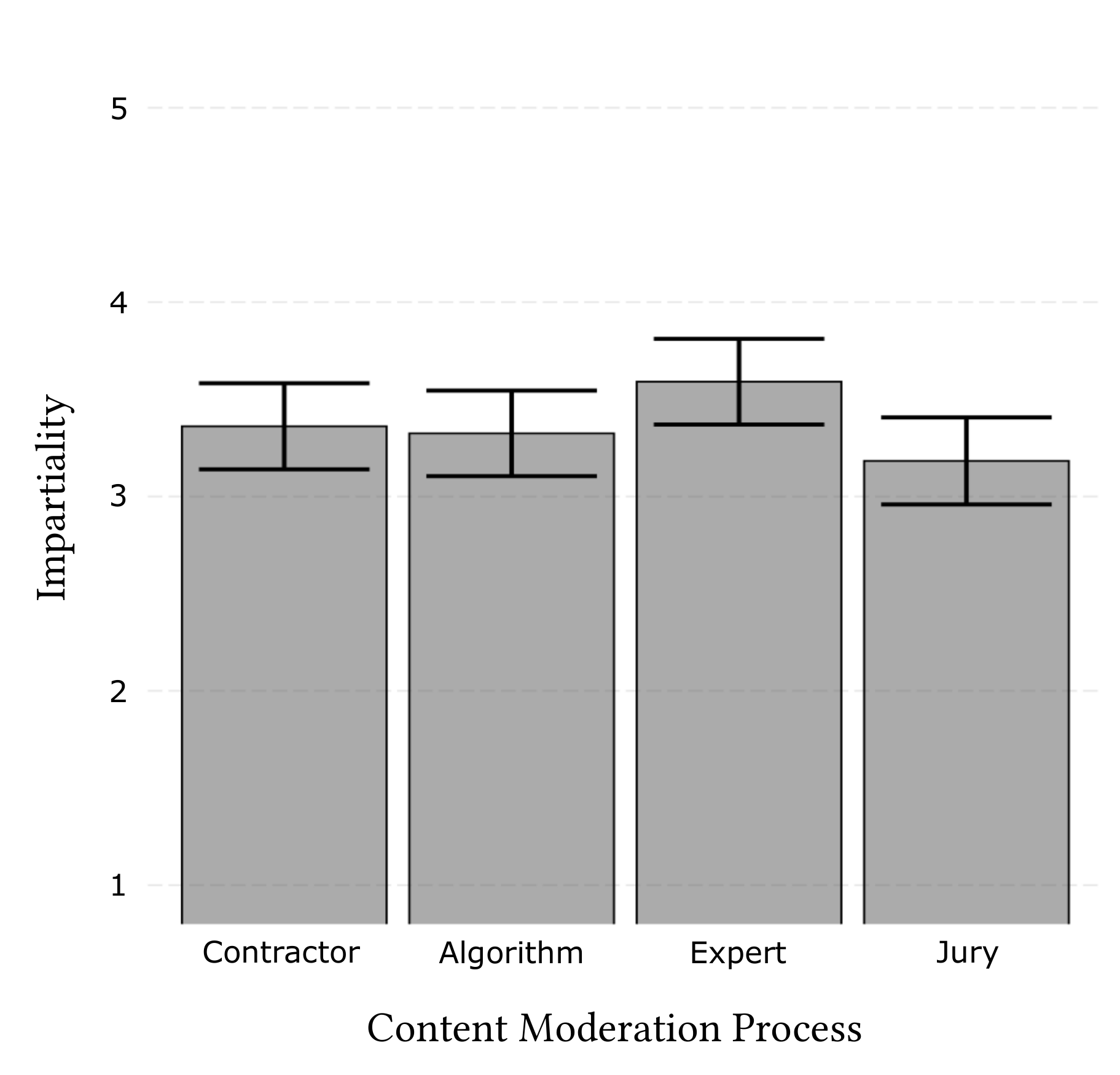}
  \caption{The mean \textit{impartiality} of each process.}
  \label{fig:impart_v_method}
\end{subfigure}
\caption{Trustworthiness and impartiality of each moderation process. Each mean is shown with a 95\% confidence interval calculated from variance of the fixed effect estimates. The Bonferroni correction used for hypothesis tests is not applied here.}
\label{fig:test}
\end{figure}

To evaluate \textbf{H1}, we conduct Tukey's HSD test as a \textit{post hoc} analysis of the pairwise differences in perceived institutional legitimacy across moderation processes. This test allows for significance testing across more than two groups and makes fewer assumptions than t-tests, which are not universally accepted for LME model parameters~\cite{luke2017evaluating}. Results are presented in Table \ref{table:h1}. From this test, we can conclude that decisions made by the expert panel are perceived as more legitimate, according to our definition, than decisions made by both the digital jury and algorithm; however there is not sufficient evidence to draw conclusions about the perceived legitimacy of other moderation processes. Consequently, we find partial support for \textbf{H1.1} and are able to disprove \textbf{H1.3},  but do not find evidence to support or disprove \textbf{H1.2}. To better visualize the varying perceived legitimacy of the four moderation processes, a marginal effects plot is presented in Figure \ref{fig:leg_v_method}. Additionally, a summary of coded free responses to the comparative questions is presented in Table \ref{table:rationales_trustworthiness} and \ref{table:rationales_impartiality}, and corresponding quantitative results are presented in Figure \ref{fig:test}.

\subsubsection{H1.1}

Quantitative results show that the expert panel has higher perceived legitimacy than the algorithm, supporting one component of \textbf{H1.1}. In free response, an important factor for participants appeared to be whether decisions were made by groups or individuals. 24\% of respondents suggested that contractors would make more biased decisions as single individuals, and 24\% suggested that contractors would apply their own beliefs and agenda. Moreover, a full 42\% of participants expressed support in some form for the idea that groups of moderators can be more trustworthy and/or impartial that single moderators. In contrast to our expectations, qualitative results cast doubt on independence as a major factor driving perceived legitimacy. Many participants acknowledged the greater independence of the two deliberative bodies---25\% of participants expressed a belief that paid contractors would carry out the agenda and biases of the platform and 16\% expressed concern that the algorithm could be programmed with platform biases, while much smaller proportions expressed similar ideas about expert panels or digital juries (2\% for both). However, only 8\% of participants selecting expert panels and 6\% of those selecting the digital jury as the most impartial process provided independence as a rationale, with similar or smaller proportions among those selecting these processes as the most trustworthy.

\begin{table}[tb]
\begin{tabular}{lllll} \toprule
\textit{Pr.} & \textit{Rationale} & \textit{\% {\small (n)}} & \textit{Rationale}   & \textit{\% {\small (n)}} \\
\midrule
 & \multicolumn{2}{c}{\textit{Highest Trustworthiness}}                 & \multicolumn{2}{c}{\textit{Lowest Trustworthiness}}                  \\
\midrule
{\footnotesize Contractor} & {\footnotesize It's their job}  & {\footnotesize 38\%} {\footnotesize (5)} & {\footnotesize Single person bias} & {\footnotesize 61\%} {\footnotesize (19)} \\
& {\footnotesize Has necessary training and knowledge} & {\footnotesize 31\%} {\footnotesize (4)} & {\footnotesize Implements platform agenda and biases }  & {\footnotesize 39\%} {\footnotesize (12)} \\
\midrule
{\footnotesize Algorithm} & {\footnotesize Decides based on logic, data, rules}   & {\footnotesize 32\%} {\footnotesize (9)}  & {\footnotesize Lacks human factors of cognition} & {\footnotesize 50\%} {\footnotesize (15)}\\
& {\footnotesize Doesn't apply own beliefs and agenda} & {\footnotesize 18\%} {\footnotesize (5)}  & {\footnotesize Generally performs poorly} & {\footnotesize 40\%} {\footnotesize (12)}  \\
\midrule
{\footnotesize Expert} & {\footnotesize Has necessary training and knowledge}   & {\footnotesize 50\%} {\footnotesize (19)}  & {\footnotesize Performed worse (in survey)} & {\footnotesize 25\%} {\footnotesize (2)} \\
& {\footnotesize Multiple people helps mitigate bias}   & {\footnotesize 26\%} {\footnotesize (10)}  & {\footnotesize Applies own beliefs and agenda} & {\footnotesize 13\%} {\footnotesize (1)} \\
\midrule
{\footnotesize Jury} & {\footnotesize Multiple people helps mitigate bias} & {\footnotesize 31\%} {\footnotesize (8)} & {\footnotesize Applies own beliefs and agenda} & {\footnotesize 54\%} {\footnotesize (13)} \\
& {\footnotesize Doesn't apply own beliefs and agenda} & {\footnotesize 12\%} {\footnotesize (3)}  & {\footnotesize Random selection process not sufficient} & {\footnotesize 38\%} {\footnotesize (9)} \\
\bottomrule
\end{tabular}
\caption{Most frequent rationales given for answer among participants selecting each process as having the highest and the lowest \textit{trustworthiness}. Proportions are given as percentages, with raw participant counts given in parentheses.}
 \vspace{-7mm}
\label{table:rationales_trustworthiness}
\end{table}

\begin{table}[tb]
\vspace{5mm}
\begin{tabular}{lllll} \toprule
\textit{Pr.} & \textit{Rationale} & \textit{\% {\small (n)}} & \textit{Rationale}   & \textit{\% {\small (n)}} \\
\midrule
& \multicolumn{2}{c}{\textit{Highest Impartiality}}                 & \multicolumn{2}{c}{\textit{Lowest Impartiality}}                  \\
\midrule
    {\footnotesize Contractor} & {\footnotesize Faithfully adheres to guidelines} & {\footnotesize 38\%} {\footnotesize (3)}& {\footnotesize Single person bias} & {\footnotesize 41\%} {\footnotesize (17)} \\
& {\footnotesize Is accountable for decisions} & {\footnotesize 25\%} {\footnotesize (2)} & {\footnotesize Applies own beliefs and agenda} & {\footnotesize 34\%} {\footnotesize (14)} \\
\midrule
{\footnotesize Algorithm} & {\footnotesize Doesn't apply own beliefs and agenda}  & {\footnotesize 52\%} {\footnotesize (24)}  & {\footnotesize Generally performs poorly} & {\footnotesize 50\%} {\footnotesize (6)} \\
& {\footnotesize Decides based on logic, data, rules} & {\footnotesize 37\%} {\footnotesize (17)}  & {\footnotesize Lacks human factors of cognition} & {\footnotesize 42\%} {\footnotesize (5)} \\
\midrule
{\footnotesize Expert} & {\footnotesize Has necessary training and knowledge}  & {\footnotesize 24\%} {\footnotesize (6)}  & {\footnotesize Applies own beliefs and agenda} & {\footnotesize 43\%} {\footnotesize (3)} \\
& {\footnotesize Multiple people helps mitigate bias}  & {\footnotesize 20\%} {\footnotesize (5)}  & {\footnotesize Unaccountable (e.g., lacks oversight)} & {\footnotesize 43\%} {\footnotesize (3)} \\
\midrule
{\footnotesize Jury} & {\footnotesize Multiple people helps mitigate bias}   & {\footnotesize 44\%} {\footnotesize (7)} & {\footnotesize Applies own beliefs and agenda} & {\footnotesize 56\%} {\footnotesize (18)} \\
& {\footnotesize Is independent of platform} & {\footnotesize 6\%} {\footnotesize (1)}  & {\footnotesize Lack necessary training and knowledge} & {\footnotesize 25\%} {\footnotesize (8)} \\
\bottomrule
\end{tabular}
\caption{Most frequent rationales given for answer among participants selecting each process as having the highest and the lowest \textit{impartiality}.  Proportions are given as percentages, with raw participant counts given in parentheses.}
 \vspace{-7mm}
\label{table:rationales_impartiality}
\end{table}

\subsubsection{H1.2}
Quantitative estimates of the perceived legitimacy of algorithms and paid contractors were not statistically distinguishable. However, qualitative analysis of free response provides more clues. Pluralities of respondents designated the paid contractor as the least trustworthy and least impartial decision maker, while a majority (51\%) chose the algorithm as the most impartial, suggesting some support for \textbf{H1.2}. The discrepancy between the quantitative estimates and free response answers for paid contractors is notable. These discrepancies might be due to estimation error and lack of statistical significance, or alternatively by substantive differences in the framing of the quantitative and free response questions. In quantitative questions, participants were asked to provide ratings in isolation, while in free response, they were asked to consider all four processes simultaneously. Moreover, in free response questions, participants were only asked to discuss the most and least trustworthy and impartial processes.

Participants were concerned with paid contractors implementing their own agenda and biases (24\%), despite the limited role of personal interpretation in contractor moderation in most platforms~\cite{trauma_floor, propublica_rules}, or implementing the biases of the platform (25\%). Interestingly, some participants viewed a paid relationship as a corrupting influence, while others viewed it as source of accountability. While a large proportion of respondents who labeled the contractor as untrustworthy (39\%) and partial (34\%) also expressed that the contractor would be subject to platform control, 10\% of participants in each case expressed concern that to contractors, moderation would be ``just a job.'' We anticipated that paid contractors would be perceived as less legitimate due to lack of clarity about their background and lack of faith in their expertise and ability to make nuanced judgments. In free response, however, these types of concerns were expressed by <5\% of respondents.

By contrast, 25\% of respondents made comments like ``The least trustworthy would likely be the algorithm due to the complex nature, nuance, and context of the human language. Algorithm[s] cannot navigate the complexities and subtleties of our communications.'' 16\% expressed awareness that algorithms can be programmed with built-in bias, suggesting that support can depend on specific details of how and why an algorithm is created. Additionally, 32\% of respondents made performance based arguments (as \textit{rationale} or \textit{qualification}) about algorithms, markedly higher than for contractors (2\%), digital juries (9\%), and expert panels (8\%). Even many participants who expressed support for algorithmic moderation had reservations. Although nearly one third of respondents believed the algorithm was the most trustworthy process, this support was made conditional at the highest rate of all processes, depending on factors like the algorithm being constructed fairly and impartially (25\%) and decisions being subject to checks and balances (11\%) and appeal to humans (7\%), with similar rates for \textit{impartiality}.

\subsubsection{H1.3}

Quantitative and qualitative results definitively refute \textbf{H1.3}, and both show a strong preference for the expert panel. We anticipated that the greater democratic legitimacy of digital juries and skepticism of claims to expertise in content moderation would override other considerations. While we did find some support for these phenomena in free response, by and large participants viewed expert panels as legitimate, trustworthy, and impartial. Although we anticipated juries' democratic nature might be seen as a check on the platform's ability to impose its own standards on the community (a view articulated by few respondents), participants seemed more concerned that digital juries would impose members' own viewpoints (expressed by 30\%) and that vetting would not be rigorous enough (26\%). One participant stated, ``It would be very difficult for users who liked a person who posts things that violated the standards to be impartial...'' Another commented, ``they are randomly chosen and could be just about anybody. If there was some type of selection process from Facebook users, then that would be a little bit different.'' One participant even fretted about demographic bias in randomly selected juries, saying ``Facebook users tend towards certain demographics -- the middle aged and not people like me who are younger.'' Additionally 8\% expressed the idea that regular users are inherently unsuited to the task. Some participants went as far as to reject the legitimacy of juries in the justice system, in one instance, stating, ``[The unfairness of juries of users] is similar to how ineffective an actual jury is at trial.'' By contrast, 25\% of participants showed appreciation for the expert panel members' training and expertise, suggesting that their perceived greater formal education and experience would help mitigate bias. 

\subsection{Importance of Outcome-Preference Alignment}

\begin{figure}[t]
    \centering
    \includegraphics[scale=0.45]{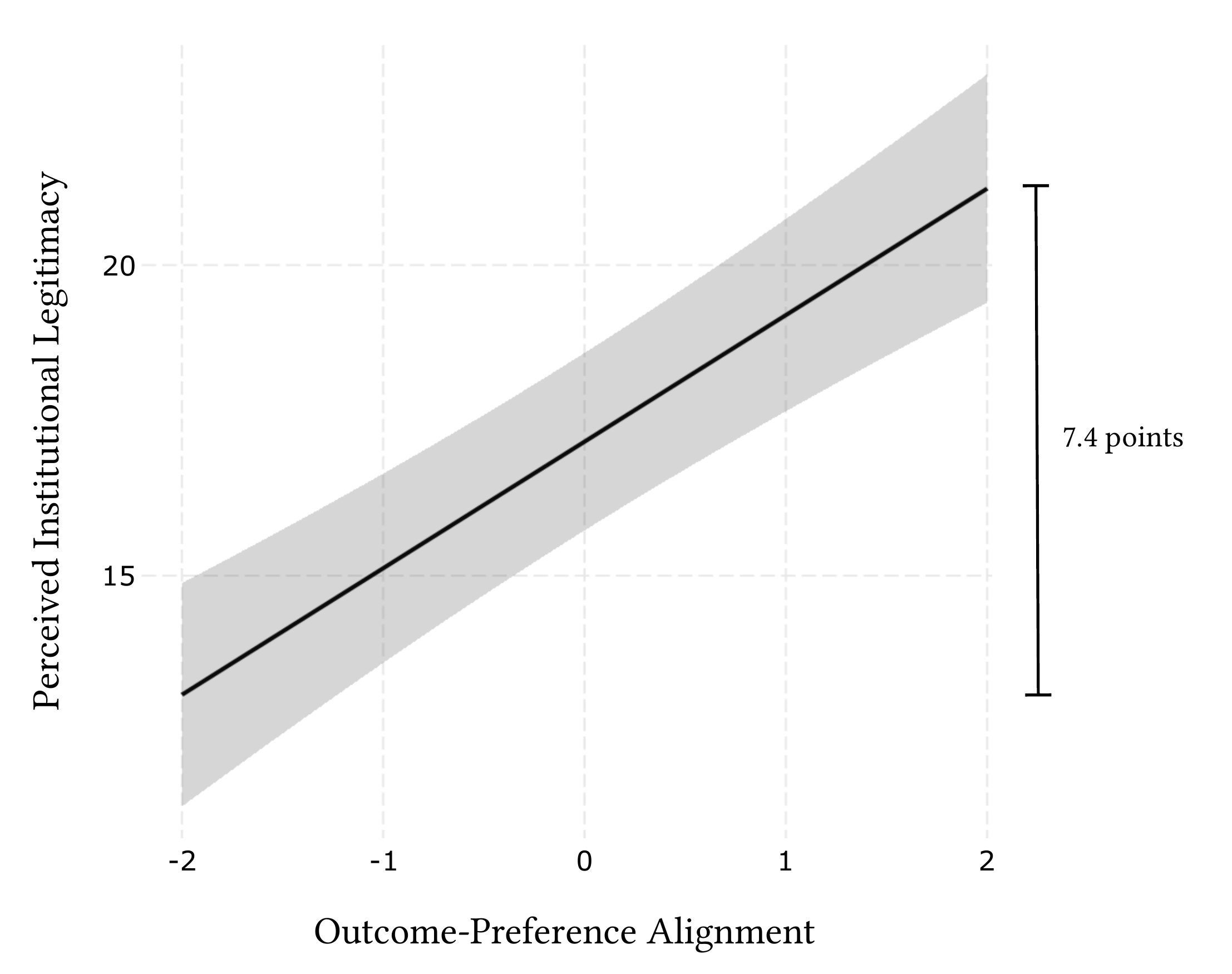}
    \caption{Shows the approximately 7.4 point range of variation (out of 20) of perceived legitimacy attributable to the linear marginal effect of \textsc{Alignment}. A 95\% confidence interval is shown calculated from variance of the fixed effect estimates.}
    \label{fig:leg_v_align}
\end{figure}

To assess \textbf{H2}, we consider both qualitative and quantitative factors. Qualitatively, the magnitude of the fixed effect of \textsc{Alignment}, as well as its significance lends support to \textbf{H2}---that users will report higher perceived legitimacy when content moderation systems make decisions that align with their individual preferences. Since \textsc{Alignment} is on a five-point scale, the maximal variation in perceived legitimacy due to \textsc{Alignment} is approximately 7.4 points out of 20, larger than that of any other variable (see Figure \ref{fig:leg_v_align}).

Quantitatively, we find that comparing the model with alternative models with a single variable removed, the largest regression occurs when removing \textsc{Alignment} according to the Akaike information criterion (AIC), an information theory based measure that balances goodness of fit with model complexity. Performing an ANOVA comparison between the primary model and a model without \textsc{Alignment}, we calculate $p<.001$. Additionally, we calculate the marginal $R^2$ value~\cite{nakagawa_2013} for a reduced model using \textsc{Alignment} as the sole predictor variable as 0.27, suggesting that 27\% of the variance in perceived legitimacy is explainable by outcome-preference alignment, assuming the modeled random effects.

Our model estimates that interactions between \textsc{Alignment} and \textsc{Process} are small and not statistically significant. Parameter estimates of these interactions are presented in Table \ref{table:regression}.

In free response, many participants' explanations for their legitimacy ratings rested solely on their opinion of the moderated post and the random decision shown to them in the study. In addition, when participants were asked to assess the overall \textit{trustworthiness} and \textit{impartiality} of moderation processes, a significant proportion made arguments based on the survey examples.

\section{Discussion}

\subsection{Implications of Process Effects}
\label{sec:discussion_methods}

Our quantitative and qualitative results build a strong case that the Expert Panel is perceived as the most legitimate process by our participants. This result might be considered surprising in light of common criticisms that platforms are undemocratic and biased in favor of unpopular views~\cite{kaye2019speech}. The result is especially notable given the limited information participants were given regarding panelist selection, ideological alignment, and the nature and relevance of their expertise. Moreover, our results' direct refutation of \textbf{H1.3} seems to show that expertise, rather than the body's independence from the platform or other characteristics, was what participants appreciated. Perhaps, despite the popular notion of a crisis of mistrust in expertise~\cite{eyal2019crisis}, mistrust of peers is stronger still~\cite{pew_trust_peers}, though different results may be obtained in high trust societies~\cite{delhey_2005} or those whose cultures are poorly represented by the expert body. As one participant noted, ``they are experts, they know how to deal with things like this better than anyone. They can be trusted more to make the right decisions.''

With respect to digital juries, other work studying online communities finds similar mistrust of peers and resistance to peer judgment as we observed~\cite{kou_2017, fan2020digital}. Digital juries might offer benefits in certain scenarios, by aligning content moderation enforcement with users' preferences. However, there remains debate around the ability of digital juries to scale effectively and carry out moderation decisions on platforms that lack diversity like Parler~\cite{parler_wsj, parler_karaswisher}.

Recent rulings by the Facebook Oversight Board, in particular its rulings about President Trump's posts following the 2021 Capitol Riot~\cite{fbTrumpRuling} show that the the exercise of expert authority in content moderation can be fraught in ways that go beyond the process factors examined in this study. Firstly, it is clear that the composition of the body, a variable not manipulated in this study, strongly colors decisions. The board, with heavy representation from lawyers and judges, has couched its reasoning within the framework of judicial review, self-imposing significant limits on the scope of its powers~\cite{neal2021}. Secondly, experts may face challenges in claiming and exercising authority, like authority to craft policy, and may be tempted to take a middle of the road approach in controversial cases to safeguard their own perceived legitimacy in the short term~\cite{douek2021}. Lastly, while limiting the body's scope may help avoid controversy, in practice it may push important work like determining how international human rights law applies to content moderation~\cite{douek2021}, to platform-internal processes with less legitimacy and transparency.

Despite inconclusive results for \textbf{H1.2}, qualitative analysis does support many phenomena regarding algorithmic decision making discussed in prior work---these phenomena suggest widespread belief in algorithmic objectivity but also show several factors limiting trust. The most widely articulated of all attitudes toward moderation processes in the free response was that algorithms don't apply their own beliefs and agenda to decisions, and the fourth most common was that algorithms make decisions based on logic and rules, not feelings. However, study results show that belief in impartiality does not necessarily translate into a high level of trust or perceived legitimacy---a similar number of respondents gave the algorithm as the least trustworthy process as had given the paid contractor. Prior work notes that positive sentiments toward algorithmic decision making are tempered by factors like the subjectivity of the domain~\cite{logg, sinha2001comparing}, opaqueness of function and deployment~\cite{eslami_2019, kizilcec_2016}, and performance~\cite{dietvorst2015algorithm}, elements that can be seen in free response.

\subsection{Implications of Outcome-Preference Alignment}

Quantitative results firmly support \textbf{H2}, showing that outcome-preference alignment strongly determines perceived institutional legitimacy. This influence far outstrips that of the process variables manipulated in this survey. This result arguably poses an intractable problem for platforms, discussed further in Section \ref{sec:design_implications}, and raises important questions about the perceived legitimacy of majoritarian decision making.

While it is difficult to disentangle outcome-preference alignment and performance in qualitative analysis, outcome favorability bias is a well documented phenomenon in both criminal justice~\cite{lind1988social} and algorithmic decision making~\cite{wang_2020}. In contrast to these contexts, however, outcome favorability for bystanders in content moderation is driven less by personal interest and more by beliefs, ideology, and community norms. We can expect, therefore, for personal experience and political discourse to be especially important in shaping presences and the subjective experience of content moderation, as discussed below.

\subsection{Familiarity, Understanding, and Experience}

Perceived legitimacy is a sociological phenomenon, and can only be meaningfully studied in the context of a society and the attitudes of individuals therein; however, it naturally follows that levels of perceived legitimacy will vary with the nature of and degree of public awareness, understanding, and idiosyncratic experience.

In comparative free response, five respondents gave a \textit{rationale} or \textit{qualification} that they did not understand a process well enough. Some respondents expressed this skepticism forcefully, for example, writing, ``...my question is, what are the experts experts in? How do we verify their expertise, and ensure they are operating in an unbiased manner?'' and ``I do not trust the algorithm because I'm unsure how it was made and what it is looking for in a content in order to determine if it should be removed or not.'' However, transparency doesn't necessarily confer trust~\cite{ananny_2018}, especially for algorithmic decision making~\cite{cheng_2019}. As users gain more understanding of the true capabilities of algorithms, they may instead grow more skeptical~\cite{de2011we}. Because content moderation today remains opaque to users~\cite{roberts2019behind, gorwa_2020}, it is important to ask the question how more knowledge might affect users' attitudes. By the same token, steps taken by platforms to help build legitimacy can only be effective when users know about them.

Personal experience can also play a major role in shaping attitudes. For algorithmic moderation, as performance improves, positive personal experiences with algorithms would be the likeliest path to changing attitudes. For unfamiliar, emerging content moderation processes, like the digital jury, initial user experiences with the system will be especially important. Although this study did not directly measure or control for familiarity or experience with content moderation mechanisms, future work can explore a single individual's subjective experience of legitimacy due to personal experience.

\subsection{Political Discourse}

Despite inconclusive quantitative results on the effect of political affiliation, it is clear that the role of political discourse in shaping attitudes toward content moderation processes cannot be ignored. Anecdotally, multiple participants complained that moderators would be chosen to reflect a political viewpoint, and one participant consistently voiced mistrust of the platform due to its liberal bias. In such cases, political affiliation appeared to play a strong role and some weak patterns emerged---experts were presumed by some users to have liberal bias, and digital juries were presumed to be more tolerant of harmful content. In the United States, content moderation has become a flashpoint, and is viewed by many conservative-leaning individuals as an illegitimate attempt to regulate speech. Rhetoric from partisan opinion leaders, for example the Republican-led FCC's announcement that it would try to reduce liability protections for platforms that moderate content~\cite{fcc_230}, both reflects and shapes public opinion. Moreover, prior work shows that reactions to hypothetical interventions taken by social media platforms can be heavily influenced by party ideologies~\cite{gron_2020}, and that more generally, liberals and conservatives place a differing degree of importance on components of perceived legitimacy, such as fairness~\cite{graham2009liberals}. In this study, qualitative responses show evidence of systemic skepticism on the part of those identifying as conservatives or independents. In any case, the prominence of content moderation as a political issue adds an element of volatility to any attempt to build legitimate moderation systems.

\subsection{Design Implications}
\label{sec:design_implications}

Although the strong effect of outcome-preference alignment appears to pose a daunting challenge for platforms, our findings suggest platforms have procedural levers at their disposal to build perceived legitimacy. We outline several such suggestions below, synthesizing our findings with analysis of prior work, industry developments, and speculation. However, we note that in general these should be implemented as part of a tiered, hybrid system that not only optimizes for perceived legitimacy but also allows for fast response times in cases where there is a likelihood of immediate harms (e.g., 2019 Christchurch terrorist attack~\cite{newyorker_fb_christchurch}), and accommodates the challenges of scale. Moreover, adopting these measures is only a first step for platforms---indeed, some are already in use. Perceived legitimacy cannot exist without both transparency and public awareness of these efforts.

Because our findings offer clear support for expert panels, we recommend that such bodies be incorporated into moderation procedures. In practice, however, it would be impractical for such panels to make a large proportion of moderation decisions. Platforms should explore alternative means to incorporate expert judgment into hybrid processes. As a first step, we suggest that a publicly visible and independent expert panel be responsible for drafting moderation guidelines. A next step would be to allow the expert panel to handle appeals of the most controversial cases~\cite{kaye_smcs}. By contrast, the Facebook Oversight Board has focused first on deciding borderline cases, and does not have the authority to set policy (though it may recommend policy changes when solicited to do so)\cite{fob_overview}. Such an appeals body is especially important when algorithmic moderation is used, given the perceived importance of oversight among study participants. However, expertise can be brought to other places. Digital juries might, for example, include an expert member to facilitate deliberation. More broadly, an independent expert group might be given authority over the overall moderation process. Finally, experts might play a visible role in training rank-and-file moderators, assessing the performance of and appropriate scope for automated systems, and educating the public about the content moderation process.

Because large proportions of our participants displayed wariness of individual moderator biases as well as groupthink in deliberative bodies, we recommend that platforms incorporate multiple perspectives into all processes. While our results might seem to imply majoritarian decision making can be seen as a legitimate in a utilitarian sense, we believe diverse perspectives are even more critical in a divided environment. Platforms should, for example, make clear to users that posts are reviewed by multiple contractors, assuaging our participants' fears that contractors apply their own biases and opinions to moderation. While Facebook is known to monitor agreement between contractors~\mbox{\cite{trauma_floor}}, our results show this is not part of the public consciousness. Deliberative bodies could also employ pre-screening to encourage more diverse composition.

To address concerns about members of digital juries applying personal biases to decisions, we suggest exploring public reputation systems to improve accountability for decisions. Our results suggest that anonymity and lack of vetting hinders accountability. Reputation systems could range from publishing jury deliberation and justifications of decisions to a numerical rating system driven by peer reviews. Similar methods could also improve the accountability of contractors.

We can also look to prior work for solutions to the outcome-preference alignment problem. It is informative to consider prior work on the US Supreme Court, an institution that is forced to make polarizing, politically charged decisions in the public eye. Gibson, for example, suggests legitimacy can arise through a social learning process~\cite{gibson2007legitimacy}. It is reasonable, therefore, to conclude that platforms may be able to improve perceived legitimacy over time through sustained public education efforts. Platforms should publish information like how automated systems are constructed and how moderators are selected and trained. Gibson also writes about the negative effects of politicization for perceived legitimacy~\cite{gibson_1992}, an outcome platforms should take care to avoid.

We can also look to the literature on procedural justice---there is evidence, for example, that perceptions of legitimacy are enhanced when authorities take extra time to explain how they reached decisions~\cite{tyler_youtube}, that having the opportunity to express views and opinions, as a user might have during an appeal, can enhance feelings of procedural fairness irrespective of outcome~\cite{lind1988social}, and that mere knowledge of such a right can have beneficial effects even if not availed~\cite{tyler_youtube}. An analogous phenomenon has also been described in the context of online content moderation~\cite{jhaver2019does}, and it is likely that enhancing the quantity and quality of communication between the user and platform can improve perceived legitimacy.

\subsection{Limitations}

While this study provides a novel comparative perspective on content moderation processes, the study design has several limitations. First, the study attempts to measure and analyze prevailing public attitudes toward content moderation processes. However, we recognize that the formation of attitudes is a multi-faceted social and experiential process, and our study design does not allow rigorous claims about attitude formation. Furthermore, our study measures attitudes at a single snapshot in time---we did not provide an opportunity for participants to gain experience with each process, instead exposing participants to a single decision per process. In addition, the study focused on highly disagreed-upon posts, and results may not generalize to all types of moderated content. Additionally, since the study only investigated one possible version of each process type, results may not generalize to all possible versions of these processes. A future study could not only examine more versions of these processes, but also identify which attributes (e.g., jury rules) contribute to perceived legitimacy.

Since our study was scoped to only include industrial content moderation, we did not investigate moderation and artisanal moderation, which gives rise to two limitations. The first is that we know less about the perceived legitimacy of the excluded approaches.  Second, this work only investigates how moderation processes impact the perceived legitimacy of \textit{rule enforcement}, not \textit{rule creation}. Thus, the results may not generalize well to the perceived legitimacy of moderation processes involved in rule creation.

Biases in our user population may also limit generalizability of results. While the study was conducted among Facebook users, the demographics of AMT workers do not exactly match that of Facebook’s US user base. The survey population overrepresented higher-educated and non-Hispanic white individuals and underrepresented multiple minority groups. Furthermore, the technical and digital nature of AMT work may mean that our survey respondents had a different relationship with online platforms than average social media users. Additionally, our study was limited to one social media platform (Facebook) in one country (United States). In addition, since attitudes and perceptions vary upon their existing knowledge of content moderation, the results may not generalize to populations with highly expert populations. A future cross cultural study may be needed to determine which drivers of perceived legitimacy are more universal and which are more specific to the United States, its present cultural moment, and the present level of knowledge about content moderation.

Additionally, two potentially significant factors of moderation processes we do not consider in this work are when moderation is carried out (i.e., pre-moderation vs. post-moderation)~\cite{veglis2014moderation, cambridge_consultants}, and tiered or hybrid moderation processes. Qualitative results suggest knowledge of oversight mechanisms and appeals processes can influence perception of legitimacy, and holistic assessment of perceived legitimacy of governance mechanisms in practice requires consideration of the entire system.

\subsection{Future Work}
Future studies should more rigorously examine tiered processes, the impact of oversight, and the appeals process. Our qualitative results indicate that including these processes may be especially significant for algorithmic moderation, where participants indicated a desire for human oversight. Additionally, future work should examine how the wide array of artisanal and community-driven moderation models found in platforms like Reddit, Vimeo, Patreon, Wikipedia, and League of Legends affect perceived legitimacy when employed together with or in place of industrial content moderation processes. Given that these approaches tend to have more community participation in governance, future work comparing them needs to be careful to separate the investigation of moderation legitimacy from that of governance.

A second area for future work is to investigate how hybrid moderation processes can better incorporate expertise. While we can hypothesize the benefit of expertise for perceived legitimacy will diminish the further removed experts are from day-to-day decision making, hybrid models are the only practical solution to scaling challenges. A promising direction might be to combine elements of the expert panel and digital jury---for example, including an expert facilitator or introducing credentialing for jury members. Additionally, future work should examine the importance of specific types of expertise and representation of diverse viewpoints.

A third critical area is to investigate the impact of political affiliation and political debate on perceived legitimacy. While quantitative and qualitative results hinted that conservatives and independents may be less trusting of content moderation processes in general, the power of the study was not sufficient to establish this. Future studies can not only investigate this effect, but also examine the interaction of political affiliation with elements of moderation process design. Moreover, while this study did establish the importance of normative preferences, it did not attempt to distinguish political or closely held preferences from other preferences, and did not specifically distinguish content with a significant political dimension from content without this dimension. An especially important topic for future work is studying how sticky factors like institutional commitment and decisional jurisdiction are in the face of politically unpalatable decisions, and what role normative concepts like democratic legitimacy play in politically charged environments.

\section{Conclusion}

As online platforms and their governance mechanisms increasingly resemble digital polities, platforms must focus greater attention on user perceptions of legitimacy. However creating a legitimate content moderation process appears to be a nearly intractable problem as long as people with different views continue to occupy the same digital spaces. Not only is the scale of the task daunting, but this study also highlights the degree to which individual outcome preferences can dominate perceptions of legitimacy, regardless of how platforms design their processes. Content for which opinions differ wildly, therefore, poses a ``catch-22'' to platforms---goodwill generated with one segment of the user population may be met in equal measure with feelings of illegitimacy by another.

Nevertheless, our quantitative and qualitative results illuminate potential paths forward. We find the strongest support for a robust role for experts in content moderation processes, with participants perceiving the expert panel as having high levels of trustworthiness, fairness and impartiality, and overall perceived institutional legitimacy. Our qualitative results also indicate a preference among users for group decision making over decisions made by individuals, supporting future work on processes that synthesize multiple views. Our results are also consistent with prior work on attitudes toward algorithmic decision making, showing that while algorithms can be perceived as legitimate decision makers, their performance and users' experience with them will significantly shape attitudes.

Today, content moderation stands at an inflection point. While platforms are accountable for their content moderation practices, academics and policymakers are increasingly vocal participants in shaping the future of content moderation, and the public will have the final say. Criticism of existing mechanisms abounds, but so do proposals and experiments seeking to build better systems. Studies of perceived legitimacy can be a powerful tool for all groups to ensure these systems are trusted and respected by the public.


\begin{acks}
We thank the reviewers for their helpful comments and suggestions. We also thank Michael Sklar of the Stanford Department of Statistics for statistical advice. The Stanford Computer Science (CS) department provided funds to assist this research. This project was supported by the Office of Naval Research (N00014-21-1-2839).
\end{acks}

\bibliographystyle{ACM-Reference-Format}
\bibliography{sample-base}

\clearpage

\section{Appendices}

\subsection{Appendix A}

\begin{table}[h]
\centering
\begin{tabular}{lp{9.5cm}}
\toprule
\textit{Content Moderation Process}   & \textit{Description} \\ 
\midrule
Paid Contractor           &      The content moderation decision was made by a human contractor employed by Facebook. The human contractor was trained in a workshop with examples of posts that violated Facebook’s Community Standards. \\
Algorithm           &     The content moderation decision was made by an algorithm that was built by software engineers at Facebook. The algorithm was trained on examples of posts that violated Facebook’s Community Standards.     \\
Digital Jury           &      The content moderation decision was made by a jury of 6 randomly-selected Facebook users. Jury members received training on enforcing Facebook’s Community Standards, and after structured deliberation in a video conference session, reached a unanimous decision.     \\
Expert Panel           &      The content moderation decision was made by a panel of 6 experts selected for their expertise in content moderation, human rights, and digital rights. After structured deliberation in a videoconference session, they reached a unanimous decision.     \\
\bottomrule
\end{tabular}
\caption{Descriptions of each of the moderation processes shown to survey participants with decision outcomes.}
\label{table:descriptions}
\end{table}

\clearpage
\subsection{Appendix B}

\begin{table}[h]
\tiny
\centering
\begin{tabular}{lp{12cm}}
\toprule
\textit{Predicate} & \textit{Description} \\ 
\midrule
1 & Subject to single person bias \\
2 & Abuses power \\
3 & Doesn't abuse power \\
4 & Applies own beliefs and agenda \\
5 & Doesn't apply own beliefs and agenda \\
6 & Allows for multiple perspectives to mitigate bias \\
7 & Suffers from group-think or peer pressure \\
8 & Has necessary formal training and experience \\
9 & Lacks necessary formal training and experience \\
10 & Takes work seriously because it's their job \\
11 & Doesn't take seriously because not compensated (enough) \\
12 & Doesn't take seriously because it's just a job \\
13 & Subject to control by the platform \\
14 & Independent from the platform \\
15 & Subject to (improper) influence by third parties \\
16 & Not subject to (improper) influence by third parties \\
17 & Has rigorous and fair selection process \\
18 & Lacks rigorous and fair selection process \\
19 & Selection controlled by the platform \\
20 & Accountable for decisions \\
21 & Unaccountable (e.g., lacks oversight) \\
22 & Makes consistent decisions \\
23 & Makes inconsistent decisions \\
24 & Faithfully adheres to moderation guidelines \\
25 & Doesn't faithfully adhere to moderation guidelines \\
26 & Makes a good-faith attempt to consider all factors and sides \\
27 & I don't understand process well enough \\
28 & Performed well in the survey examples \\
29 & Performed poorly in the survey examples \\
30 & Generally performs well \\
31 & Generally performs poorly \\
32 & Makes decisions based on logic, data, and/or rules, not feelings \\
33 & May have relationship with defendant \\
34 & No relationship to defendant \\
35 & Has human factors of cognition \\
36 & Lacks human factors of cognition \\
37 & Can be programmed with biases \\
38 & Can be trained or programmed poorly \\
39 & Can be optimized or improved over time \\
40 & Costly or impractical \\
41 & Composed of regular users (who are well equipped to make decisions) \\
42 & Composed of regular users (who are unsuited to make decisions) \\
43 & I do not trust \\
44 & Resembles criminal justice system \\
45 & I trust \\
46 & Considers broader social context \\
47 & Does not consider broader social context \\
48 & Concerned with upholding individual rights \\
\bottomrule
\end{tabular}
\caption{Predicates used for qualitative coding of the free responses written by survey participants. Each predicate represents a core idea expressed in the participant’s response, and can be used to represent a rationale or a qualification for the opinion expressed.}
\label{table:predicates}
\end{table}

\clearpage
\subsection{Appendix C}

\begin{table}[h]\centering
\footnotesize
\begin{tabular*}{\columnwidth}{@{\extracolsep{\fill}} llllll }
\toprule
\textit{Variable}   & \textit{Satisfaction} & \textit{Impartiality} & \textit{Trustworthiness} & \textit{Commitment}  & \textit{Jurisdiction} \\ 
\midrule
Alignment           &\textbf{0.66***} &\textbf{0.29***} &\textbf{0.27***} &\textbf{0.34***} &\textbf{0.33***} \\
                    &      (0.12)     &      (0.03)     &       (0.03)    &       (0.04)    &       (0.04)    \\
Algorithm           &      -0.01      &\textbf{-0.18*}  &       -0.07     &       -0.14     &       -0.11     \\
                    &      (0.03)     &      (0.08)     &       (0.08)    &       (0.09)    &       (0.09)    \\
Expert Panel        &       0.07      &\textbf{0.22**}  &\textbf{0.21**}  &\textbf{0.25**}  &\textbf{0.36***} \\
                    &      (0.08)     &      (0.08)     &       (0.08)    &       (0.09)    &       (0.09)    \\
Digital Jury        &      -0.03      &      -0.04      &\textbf{-0.18*}  &       -0.18     &\textbf{-0.27**} \\
                    &      (0.08)     &      (0.08)     &       (0.08)    &       (0.09)    &       (0.09)    \\
Male                &      -0.03      &      -0.09      &       -0.06     &       -0.14     &       -0.09     \\
                    &      (0.13)     &      (0.15)     &       (0.16)    &       (0.15)    &       (0.17)    \\
Conservative        &      -0.22      &      -0.24      &       -0.32     &       -0.33     &       -0.02     \\
                    &      (0.17)     &      (0.19)     &       (0.21)    &       (0.18)    &       (0.21)    \\
Independent         &\textbf{-0.37*}  &\textbf{-0.38*}  &\textbf{-0.67**} &\textbf{-0.46*}  &       -0.34     \\
                    &     (0.18)      &      (0.18)     &       (0.20)    &       (0.19)    &       (0.20)    \\
Unreported Affiliation&\textbf{-1.13*}&      -0.38      &       -0.90     &       -0.92     &       -0.38     \\
                    &     (0.45)      &      (0.52)     &       (0.56)    &       (0.52)    &       (0.63)    \\
Alignment * Algorithm&     -0.05      &       0.09      &        0.02     &        0.06     &        0.04     \\
                    &      (0.06)     &      (0.06)     &       (0.06)    &       (0.06)    &       (0.06)    \\
Alignment * Expert Panel&   0.07      &      -0.03      &        0.04     &        0.01     &        0.07     \\
                    &      (0.06)     &      (0.06)     &       (0.06)    &       (0.07)    &       (0.06)    \\
Alignment * Digital Jury&   0.01      &       0.03      &        0.02     &        0.04     &       -0.02     \\
                    &      (0.06)     &      (0.06)     &       (0.06)    &       (0.07)    &       (0.07)    \\
Constant            &\textbf{3.59***} &\textbf{3.61***} &\textbf{3.65***} &\textbf{3.61***} &\textbf{3.25***} \\
                    &      (0.12)     &      (0.14)     &       (0.15)    &       (0.14)    &       (0.16)    \\
\bottomrule
\end{tabular*}
\caption{Regression coefficients of perceived institutional legitimacy submodels, whose dependent variables are outcome satisfaction, fairness and impartiality, trustworthiness, institutional commitment, and decisional jurisdiction, respectively. Process is modeled using deviation contrasts such that the Constant reflects the mean across processes, and coefficients are comparable between models. Significance levels are indicated for readability purposes only and are calculated with t-tests using Satterthwaite's method~\cite{luke2017evaluating}. These levels are not calculated with the Bonferroni correction used for hypothesis tests, and do not constitute formal hypothesis tests. | $p<0.001$\textsuperscript{***}, 
  $p<0.01$\textsuperscript{**}, 
  $p<0.05$\textsuperscript{*}}
 \vspace{-7mm}
\label{table:regression_submodels}
\end{table}

\received{January 2021}
\received[revised]{July 2021}
\received[accepted]{November 2021}

\appendix

\end{document}



\section*{Survey Screenshots}

This section contains screenshots of an example of a survey shown to participants.

\begin{figure}[h!]
    \centering
    \includegraphics[width=15cm]{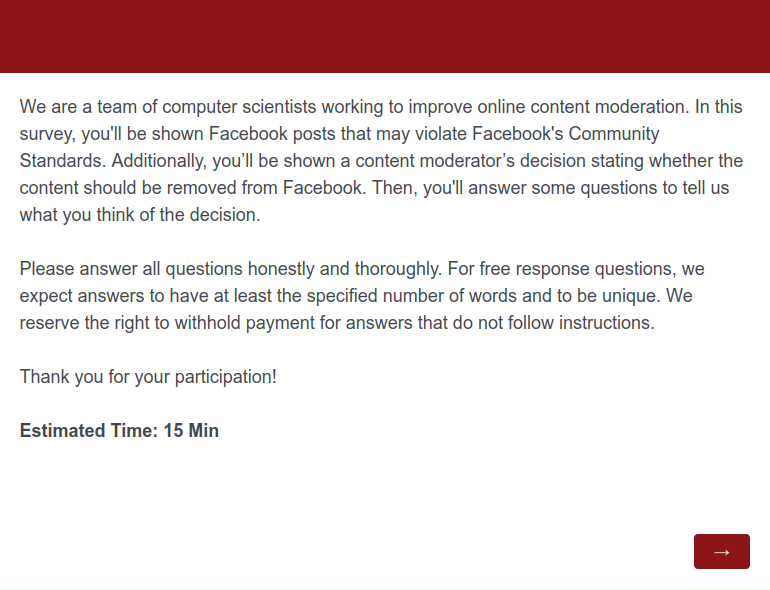}
    \caption{Survey Introduction.}
    \label{fig:survey_intro}
\end{figure}

\begin{figure}[h!]
    \centering
    \includegraphics[width=15cm]{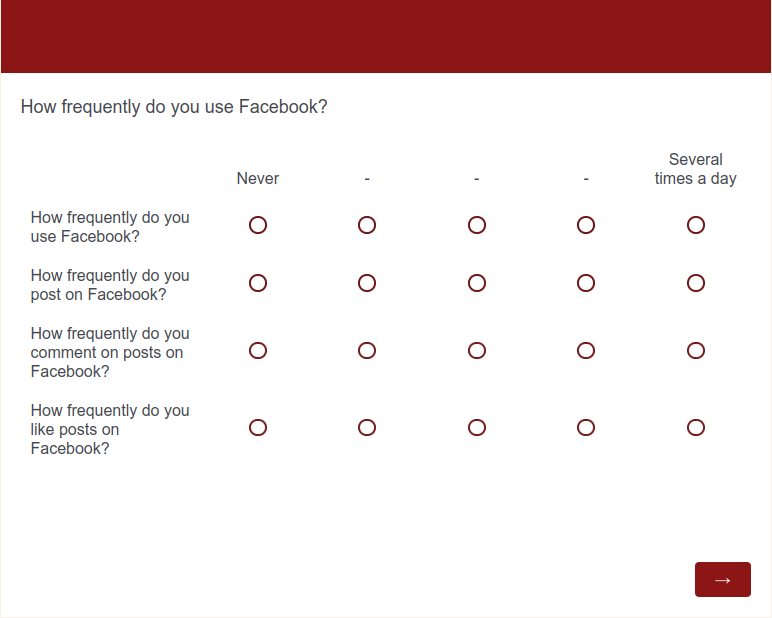}
    \caption{Participants were first surveyed on their social media usage.}
    \label{fig:survey_usage}
\end{figure}

\begin{figure}[h!]
    \centering
    \includegraphics[height=21cm]{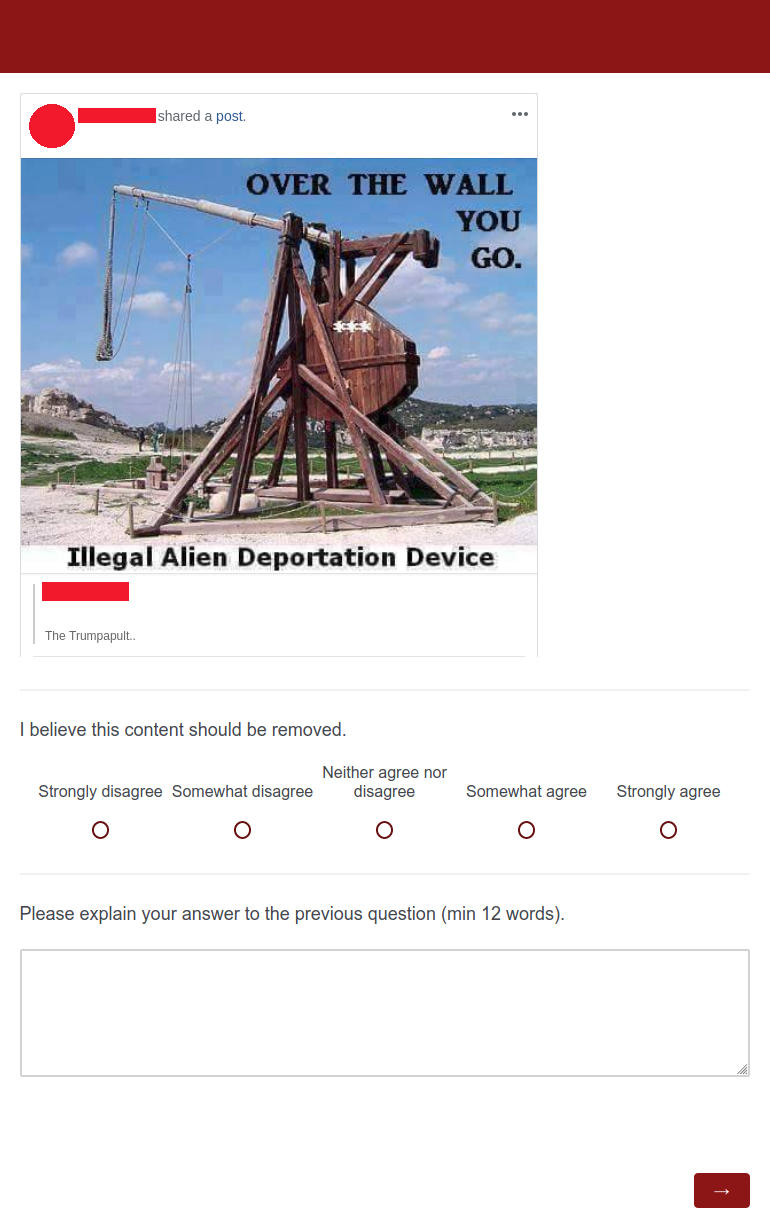}
    \caption{For each post shown to participants, participants were asked about their preferences for the moderation decision before seeing the decision.}
    \label{fig:survey_post}
\end{figure}

\begin{figure}[h!]
    \centering
    \includegraphics[height=21cm]{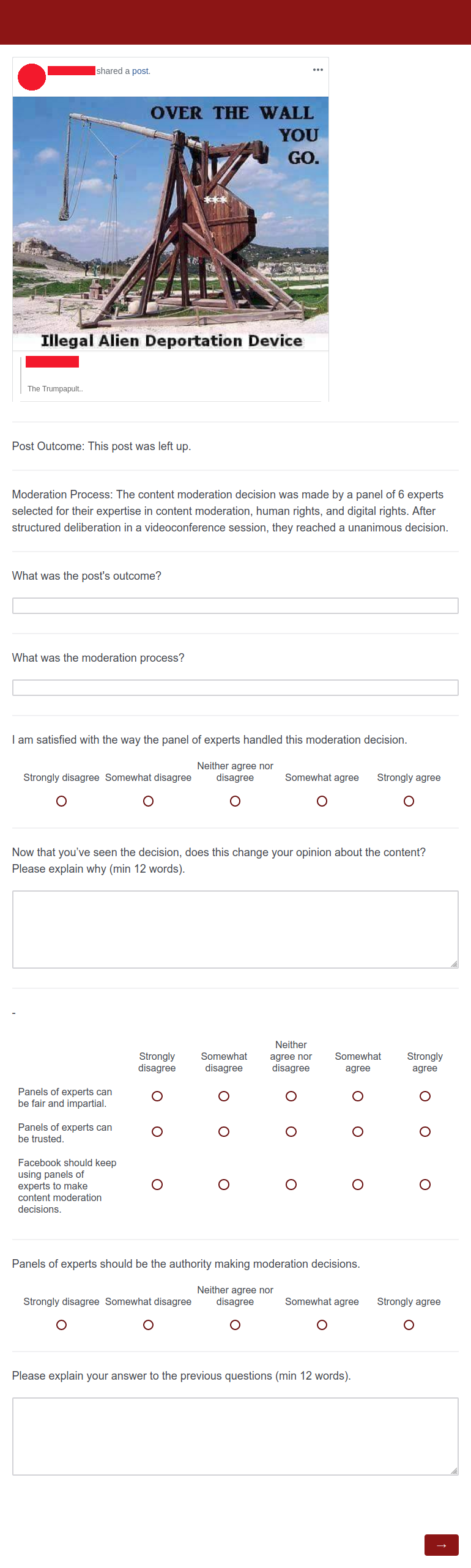}
    \caption{After each post was introduced, a random decision was shown to participants and participants were surveyed on their attitudes toward that decision.}
    \label{fig:survey_post_outcome}
\end{figure}

\begin{figure}[h!]
    \centering
    \includegraphics[height=21cm]{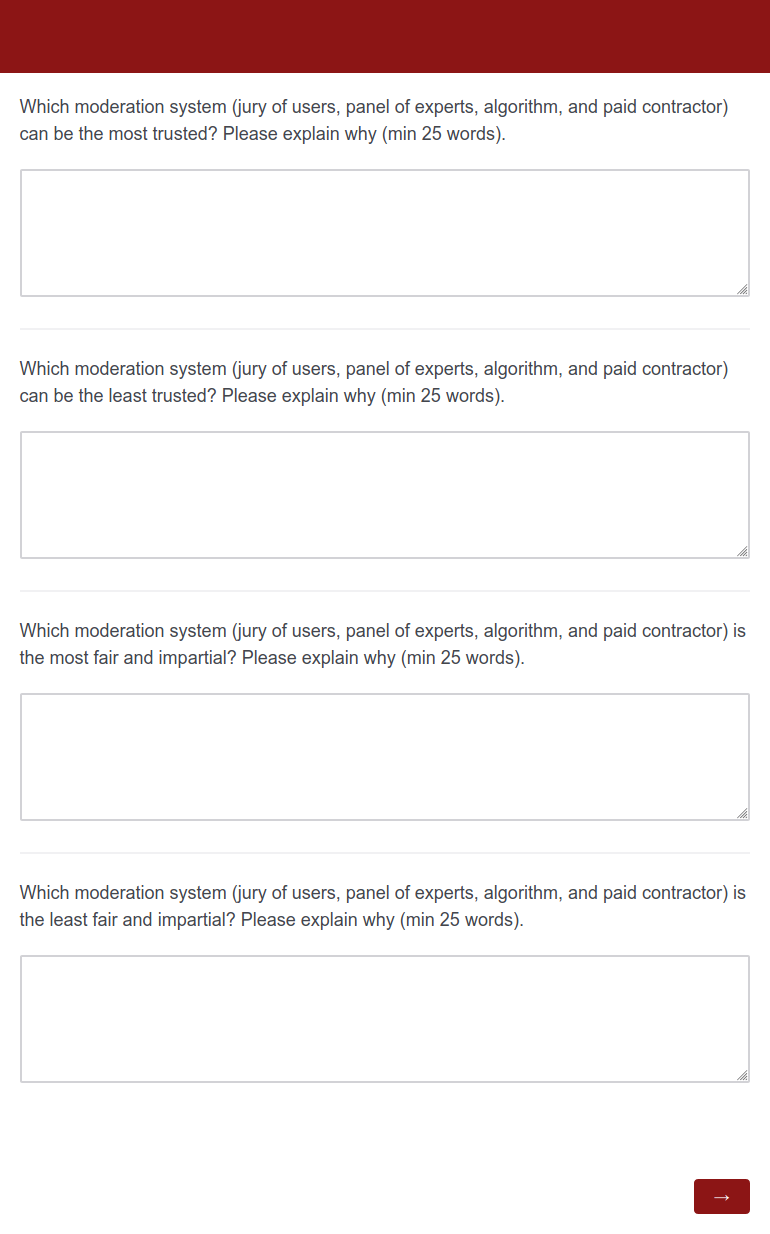}
    \caption{After participants were exposed to four posts and decisions (one for each process), they were asked to reflect on the processes on a comparative basis.}
    \label{fig:survey_post_comparative}
\end{figure}

\clearpage
\section*{Controversial Posts}

\section*{Pre-Survey}

The posts shown in the main survey (shown above) were selected among a broader pool of posts within three categories of potential violation.

\subsection*{Pre-Survey}

\subsubsection*{Hate Speech}

The authors sourced a pool of posts from public sources that may be viewed as violating the Facebook Community Standards's publicly available hate speech rules:  “We define hate speech as a direct attack on people based on what we call protected characteristics — race, ethnicity, national origin, religious affiliation, sexual orientation, caste, sex, gender, gender identity, and serious disease or disability. [...] We define attack as violent or dehumanizing speech, statements of inferiority, or calls for exclusion or segregation.” (\textbf{Source: } https://transparency.fb.com/policies/community-standards/hate-speech/). Care was taken to include a diverse set of examples, including posts directed against both majority and minority identity groups.

The following three posts, used in the main survey, were selected quantitatively from the larger pool as the most disagreed-upon posts using the methodology described in the main paper. These posts may or may not violate the Facebook Community Standards and may or may not meet the standard of hate speech according to various legal, academic, and industry definitions. Neither these three posts nor their inclusion in this survey reflect the opinions of the authors.

\begin{figure}[h!]
    \centering
    \includegraphics[width=10cm]{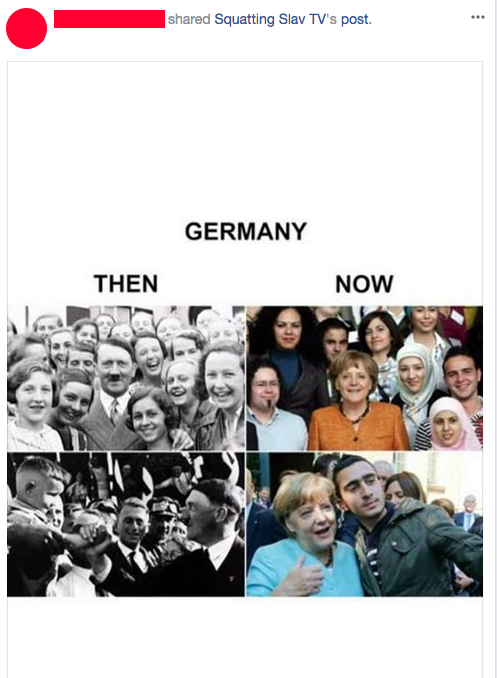}
    \caption{Post 1 (Hate Speech). This post was sampled for being highly disagreed-upon in the pre-survey as to whether it constituted hate speech that should be removed.}
    \label{fig:post1}
\end{figure}

\begin{figure}[h!]
    \centering
    \includegraphics[width=10cm]{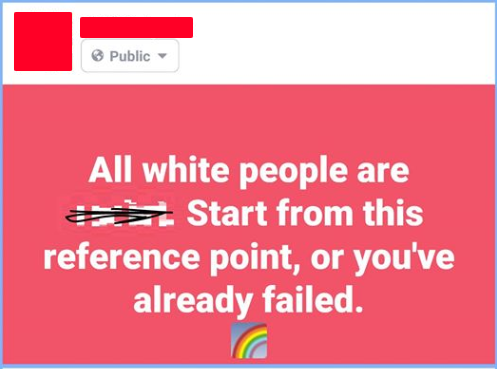}
    \caption{Post 2 (Hate Speech). This post was sampled for being highly disagreed-upon in the pre-survey as to whether it constituted hate speech that should be removed.}
    \label{fig:post2}
\end{figure}

\begin{figure}[h!]
    \centering
    \includegraphics[width=10cm]{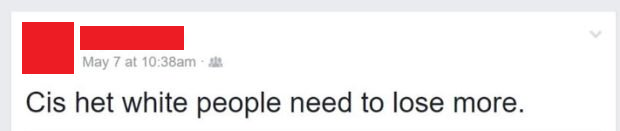}
    \caption{Post 3 (Hate Speech). This post was sampled for being highly disagreed-upon in the pre-survey as to whether it constituted hate speech that should be removed.}
    \label{fig:post3}
\end{figure}

\clearpage
\subsubsection*{Incitement to Violence}

The authors sourced a pool of posts from public sources that may be viewed as violating the Facebook Community Standards's publicly available rules pertaining to violence and incitement, prohibiting: 1) "Statements of intent to commit violence" 2) "Calls for [...] violence including content where no target is specified but a symbol represents the target" 3) "Statements advocating violence" 4) "Aspirational or conditional statements to commit violence." (\textbf{Source: }\\ https://transparency.fb.com/policies/community-standards/violence-incitement/). Care was taken to include a diverse set of examples, including posts representing multiple political viewpoints.

Two of the following three posts, used in the main survey, were selected quantitatively from the larger pool as the most disagreed-upon posts using the methodology described in the main paper. One post was selected in error, as described in the footnote in Section 3.1 of the paper. These three posts may or may not violate the Facebook Community Standards and may or may not meet the standard set by Facebook for removal or other interventions. Neither these three posts nor their inclusion in this survey reflect the opinions of the authors.

\begin{figure}[h!]
    \centering
    \includegraphics[width=10cm]{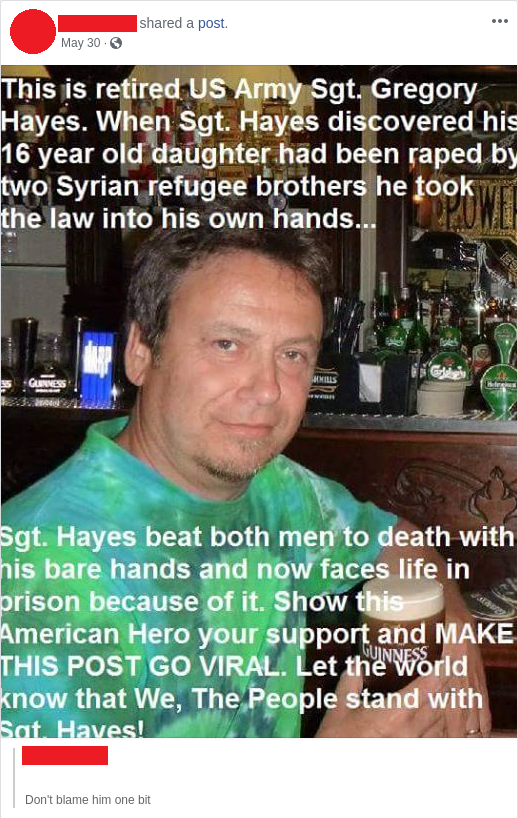}
    \caption{Post 4 (Incitement to Violence).  This post was sampled for being highly disagreed-upon in the pre-survey as to whether it constituted incitement to violence that should be removed.}
    \label{fig:post4}
\end{figure}

\begin{figure}[h!]
    \centering
    \includegraphics[width=10cm]{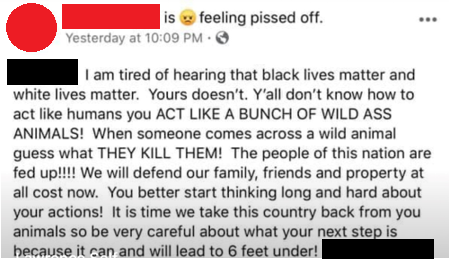}
    \caption{Post 5 (Incitement to Violence). This post was sampled in error, as explained in the footnote in Section 3.1 of the paper. However, this post was among the broader pool for the incitement to violence category.}
    \label{fig:post5}
\end{figure}

\begin{figure}[h!]
    \centering
    \includegraphics[width=10cm]{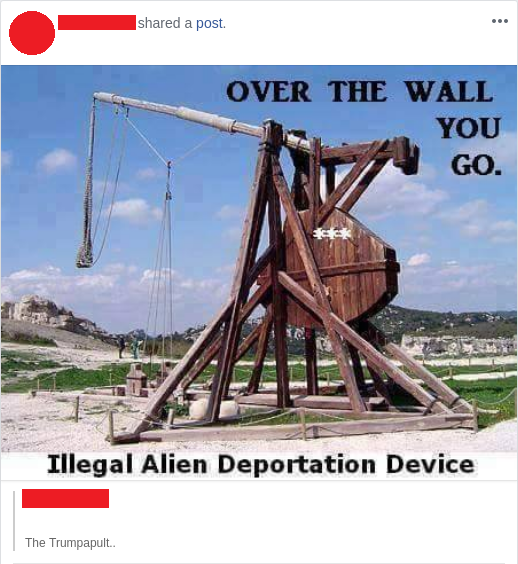}
    \caption{Post 6 (Incitement to Violence). This post was sampled for being highly disagreed-upon in the pre-survey as to whether it constituted incitement to violence that should be removed.}
    \label{fig:post6}
\end{figure}

\clearpage
\subsubsection*{Misinformation}

The authors sourced a pool of posts from public sources that may be viewed as violating the Facebook Community Standards's publicly available rules pertaining to false news and misinformation (\textbf{Sources: } https://transparency.fb.com/policies/community-standards/false-news/,\\ https://transparency.fb.com/features/approach-to-misinformation/). Care was taken to include a diverse set of examples, including posts representing multiple political viewpoints.

The following three posts, used in the main survey, were selected quantitatively from the larger pool as the most disagreed-upon posts using the methodology described in the main paper. These posts may or may not violate the Facebook Community Standards and may or may not meet the standard set by Facebook for removal or other interventions. Neither these three posts nor their inclusion in this survey reflect the opinions of the authors.

\begin{figure}[h!]
    \centering
    \includegraphics[width=10cm]{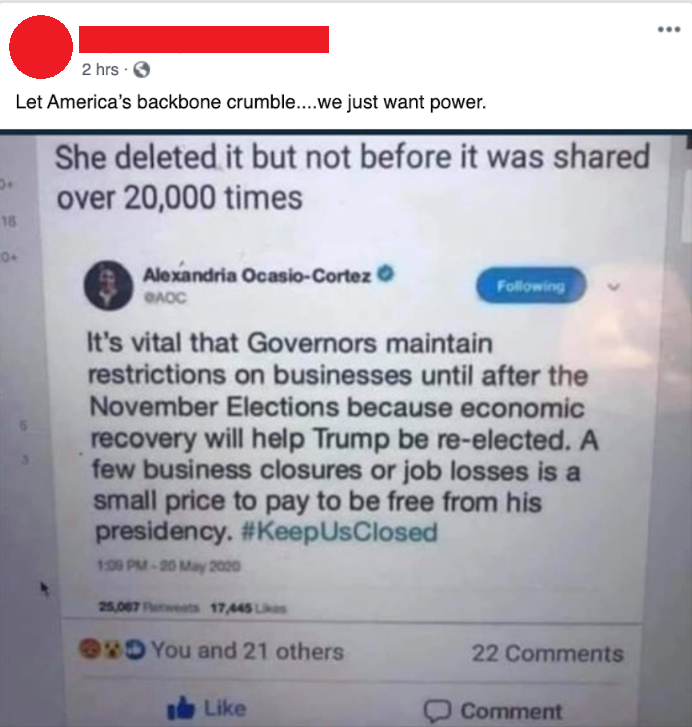}
    \caption{Post 7 (Misinformation).  This post was sampled for being highly disagreed-upon in the pre-survey as to whether it constituted misinformation that should be removed.}
    \label{fig:post7}
\end{figure}

\begin{figure}[h!]
    \centering
    \includegraphics[width=10cm]{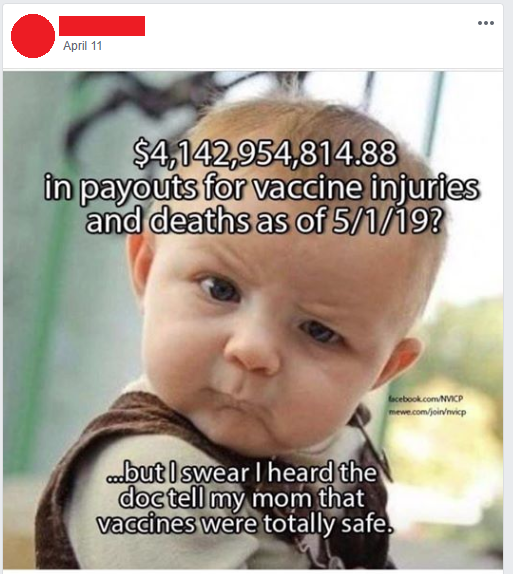}
    \caption{Post 8 (Misinformation). This post was sampled for being highly disagreed-upon in the pre-survey as to whether it constituted misinformation that should be removed.}
    \label{fig:post8}
\end{figure}

\begin{figure}[h!]
    \centering
    \includegraphics[width=10cm]{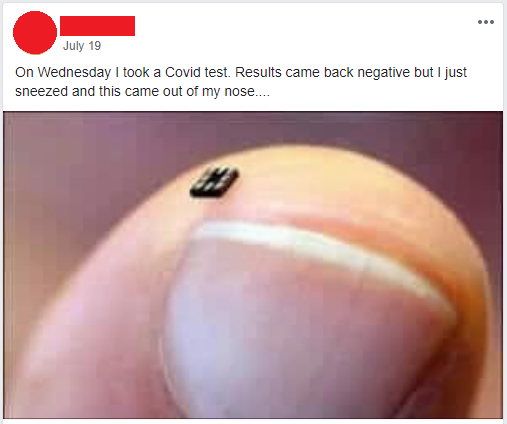}
    \caption{Post 9 (Misinformation).  This post was sampled for being highly disagreed-upon in the pre-survey as to whether it constituted misinformation that should be removed.}
    \label{fig:post9}
\end{figure}

\appendix